\documentclass[fleqn,usenatbib]{mnras}

\usepackage{newtxtext,newtxmath}

\usepackage[T1]{fontenc}

\usepackage{graphicx}	
\usepackage{amsmath}	
\usepackage{comment}
\usepackage[printonlyused]{acronym}
\usepackage{ulem}
\usepackage{threeparttable}

\newcommand{\Msun}{\mbox{M$_{\odot}$}}
\newcommand{\Rsun}{\mbox{R$_{\odot}$}}

\newcommand{\M}[1]{\mbox{$M_\mathrm{#1}$}}
\newcommand{\Mrem}[2]{\mbox{$M_\mathrm{rem {#1}{#2}}$}}
\newcommand{\Mz}[2]{\mbox{$M_\mathrm{ZAMS {#1}{#2}}$}}
\newcommand{\sevn}{\textsc{sevn}}

\defcitealias{giacobbo2020}{GM20}
\defcitealias{kroupa2001}{K01}
\defcitealias{larson1998}{L98}
\defcitealias{sana2012}{S12}
\defcitealias{stacy2013}{SB13}
\defcitealias{hartwig2022}{H22}
\defcitealias{jaacks2019}{J19}
\defcitealias{liu_bromm_20}{LB20}
\defcitealias{skinner2020}{SW20}
\defcitealias{Park2023}{P23}

\acrodef{AGB}{asymptotic giant branch}
\acrodef{EAGB}{early-asymptotic giant branch}
\acrodef{MS}{main sequence}
\acrodef{ZAMS}{zero-age main sequence}
\acrodef{CHeB}{core helium burning}
\acrodef{CC}{core collapse}
\acrodef{RSG}{red super-giant}
\acrodef{BSG}{blue super-giant}
\acrodef{YSG}{yellow super-giant}

\acrodef{PI}{pair instability}
\acrodef{PPI}{pulsational pair--instability}
\acrodef{PISN}{pair instability supernova}
\acrodef{IMF}{initial mass function}

\acrodef{BH}{black hole}
\acrodef{SSE}{single stellar evolution}
\acrodef{SBH}{single black holes}
\acrodef{BBH}{binary black holes}

\newcommand{\orcidicon}[1]{\href{https://orcid.org/#1}{\includegraphics[width=11pt]{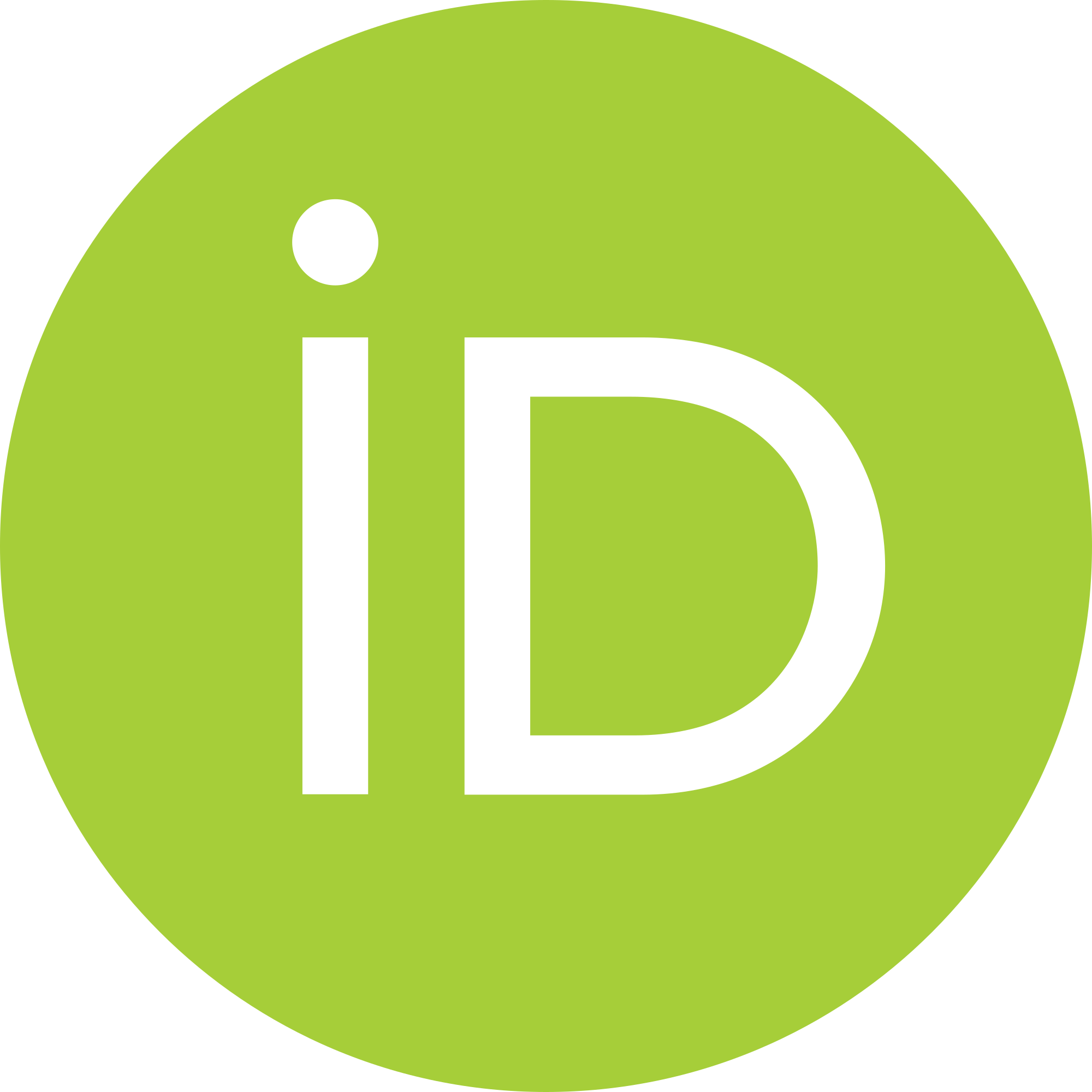}}}

\newcommand{\orcid}[1]{\href{https://orcid.org/#1}{\protect\orcidicon{#1}}}

\title[BBHs from Pop.~II and III stars]{Massive binary black holes from Population~II and III stars}

\author[Costa et al.]{
Guglielmo Costa\orcid{0000-0002-6213-6988}$^{1,2,3,4}$ \thanks{E-mail: guglielmo.costa.astro@gmail.com},
Michela Mapelli\orcid{0000-0001-8799-2548}$^{1,2,3,5}$\thanks{E-mail:michela.mapelli@unipd.it},
 Giuliano Iorio\orcid{0000-0003-0293-503X}$^{1,2,3}$\thanks{E-mail: giuliano.iorio.astro@gmail.com},
 Filippo Santoliquido\orcid{0000-0003-3752-1400}$^{1,2}$, 
 \newauthor{Gast\'on J. Escobar\orcid{0000-0002-4007-7585}$^{1,2}$, Ralf S. Klessen\orcid{0000-0002-0560-3172}$^{5}$, 
 and Alessandro Bressan\orcid{0000-0002-7922-8440}$^{6,3}$
 }
\\
$^{1}$Dipartimento di Fisica e Astronomia Galileo Galilei, Universit\`a di  Padova, Vicolo dell'Osservatorio 3, I--35122 Padova, Italy\\
$^{2}$INFN-Padova, Via Marzolo 8, I--35131 Padova, Italy\\
$^{3}$INAF-Padova, Vicolo dell'Osservatorio 5, I--35122 Padova, Italy\\
$^{4}$Univ Lyon, Univ Lyon1, Ens de Lyon, CNRS, Centre de Recherche Astrophysique de Lyon UMR5574,  F-69230 Saint-Genis-Laval, France\\
$^{5}$Universit\"at Heidelberg, Zentrum f\"ur Astronomie, Institut f\"ur Theoretische Astrophysik, Albert-Ueberle-Str. 2, D--69120 Heidelberg, Germany \\
$^{6}$SISSA, via Bonomea 365, I--34136 Trieste, Italy
}

\date{Accepted XXX. Received YYY; in original form ZZZ}

\pubyear{2023}

\begin{document}
\label{firstpage}
\pagerange{\pageref{firstpage}--\pageref{lastpage}}
\maketitle

\begin{abstract}
Population~III stars, born from the primordial gas in the Universe, lose a negligible fraction of their mass via stellar winds and possibly follow a top-heavy mass function. Hence, they have often been regarded as the ideal progenitors of massive black holes (BHs), even above the pair instability mass gap. Here, we evolve a large set of Population~III binary stars (metallicity $Z=10^{-11}$) with our population-synthesis code \sevn{}, and  compare them with Population~II binary stars ($Z=10^{-4}$). In our models, the lower edge of the pair-instability mass gap corresponds to a BH mass of $\approx{86}$ ($\approx{91}$) M$_\odot$ for single Population~III (II) stars. Overall, we find only mild differences between the properties of binary BHs (BBHs) born from Population~III and~II stars, especially if we adopt the same initial mass function and initial orbital properties. Most BBH mergers born from Population~III and II stars have primary BH mass below the pair-instability gap, and the maximum secondary BH mass is $<50$ M$_\odot$. Only up to $\approx{3.3}$\% ($\approx{0.09}$\%) BBH mergers from Population~III (II) progenitors have primary mass above the gap. Unlike metal-rich binary stars, the main formation channel of BBH mergers from  Population~III and II stars involves only stable mass transfer episodes in our fiducial model. 
\end{abstract}

\begin{keywords}
black hole physics -- stars: Population II -- stars: Population III -- gravitational waves -- methods: numerical
\end{keywords}



\section{Introduction}
Population~III (hereafter, Pop.~III) stars formed from metal-free primordial gas in the early Universe, and have eluded any attempt to observe them to date \citep[e.g.,][for a review]{bromm2004, klessen2023}.
Their initial mass function (IMF) is commonly considered to be more top-heavy than that of metal-rich stars,  mostly because molecular hydrogen is  an inefficient coolant with respect to dust \citep[e.g.,][]{bromm2004,schneider2006,stacy2013,susa2014,hirano2014, hirano2015,wollenberg2020,chon2021,tanikawa2021,jaura2022,prole2022,Park2023}. Massive  Pop.~III stars lose a negligible fraction of their mass during their life, because  stellar winds are highly inefficient for a nearly metal-free chemical composition \citep[e.g.,][]{woosley2002,volpato2023}. 
If Pop.~III stars avoid pair instability \citep{woosley2017}, they might thus end their life with a direct collapse, leading to the formation of massive black holes \citep[BHs, e.g.,][]{woosley2002}. For this reason, Pop.~III stars have been extensively studied \citep[e.g.,][]{Kinugawa2014,belczynski2017,Kinugawa2020,tanikawa2021,tanikawa2022} as possible progenitors of the most massive BHs observed by the LIGO--Virgo--KAGRA (LVK) collaboration \citep[][]{abbottGW190521,abbottGW190521astro,abbottGWTC2.1,abbottGWTC3}.

Population~II (hereafter, Pop.~II) stars formed from material that was already enriched in metals by Pop.~III stars. With a metallicity\footnote{Here and in the following, we define $Z$ as the mass fraction of elements heavier than helium, in absolute values.} ranging from $Z\sim{10^{-6}}$ to a few $\times{}10^{-4}$  Pop.~II stars are way more common in the Universe than Pop.~III stars \citep[e.g.,][]{smith2015}: we observe them in  metal-poor globular clusters, as well as in the halo of the Milky Way and in some metal-poor dwarf galaxies \citep[e.g.,][]{frebel2007,frebel2015}. It is still unclear whether Pop.~II stars follow the same IMF as metal-rich stars \citep[e.g.,][]{schneider2012,chiaki2018,chon2021,sharda2022}. 
Their metal content  is still sufficiently low that stellar winds are heavily quenched in Pop.~II stars, too \citep[e.g.,][]{chen2015}. Thus, massive Pop.~II stars might also collapse leaving massive compact remnants at the end of their life, but their contribution to the population of BHs and intermediate-mass BHs has been less investigated than that of Pop.~III stars \citep[][]{spera2017,renzo2020}.

Both massive Pop.~III and Pop.~II stars are expected to undergo pair instability or pulsational pair instability if their central temperature and density lead to an efficient production of electron and positron pairs \citep{fowler1964,barkat1967,rakavy1967,woosley2007}. If the helium core mass grows to $\sim{64-135}$ M$_\odot$ at the end of carbon burning, the star is expected to be completely disrupted by a pair instability supernova, leaving no compact remnant, while  higher He-core masses enable the direct collapse of the star to a BH \citep{woosley2002}. For He-core masses in the range $\sim{32-64}$ M$_\odot$ \citep{woosley2017,woosley2019,marchant2019,farmer2019}, pair instability triggers pulsations of the star, which enhance mass loss and, in the end, allow the star to find a stable configuration. While the boundaries of pair instability and the final compact remnant masses are still highly uncertain \citep[e.g.,][]{leung2019,farmer2019,farmer2020,marchant2019,stevenson2019,renzo2020,marchant2021,mapelli2020,costa2021,woosley2021,vink2021}, this process has a key impact on the final population of binary BHs (BBHs) born from metal-free and metal-poor stars.

Here, we model a population of BHs and BBHs born from Pop.~III and Pop.~II stars. Our Pop.~III (II) star models assume a metallicity $Z=10^{-11}$ ($Z=10^{-4}$). We probe a large range of initial configurations for the IMF and orbital parameters of Pop.~III and II binary stars. We show that the differences between the two BH populations are subtle. Both Pop.~II and III stars can give birth to very massive BHs above the pair instability mass gap. However, most BBH mergers born via isolated binary evolution host BHs below the pair-instability mass gap. When the initial semi-major axis distribution is skewed toward small values ($<10^3$ R$_\odot$), the vast majority of BBH mergers originate from Pop.~III and Pop.~II binary stars that evolve only via stable mass transfer, without common envelope. 
In a companion paper \citep{Santoliquido2023}, we explore the impact of these models on the cosmic merger rate of BBHs.

This paper is structured as follows. In Section~\ref{sec:methods} we describe our stellar tracks and population~synthesis simulations. Section~\ref{sec:results} summarizes our main results, that we discuss in Section~\ref{sec:disc} by considering the main formation channels of the simulated BBH mergers. We draw our main conclusions in Section~\ref{sec:summary}.

\section{Methods}
\label{sec:methods}

\subsection{Binary population synthesis code (\sevn{}) }
In this work, we use the \sevn{} code version 2 \citep[][]{Iorio2023}. \sevn{} integrates the evolution of  stellar properties (e.g., total mass, photospheric radius, luminosity, helium and carbon-oxygen core mass and radius) by interpolating a set of stellar tracks \citep{spera2017}, and models the main binary evolution processes (mass transfer via stellar winds, Roche lobe overflow, common envelope evolution, tides, and gravitational-wave decay) according to the semi-analytic formalism presented in \cite{hurley2002}, with several updates  described in \cite{Iorio2023}. In the following, we adopt the same set up as the fiducial model presented in \cite{Iorio2023}. In particular,  Roche-lobe overflow mass transfer is always stable for \ac{MS} and Hertzsprung gap donor stars, while we follow the prescriptions by \cite{hurley2002}  in all the other cases. We model the mass accretion rate during Roche-lobe overflow as
\begin{equation}
 \dot{M}_\mathrm{a} = \left\{
 \begin{array}{ll}
 \min{(\dot{M}_{\rm Edd},\,{}-f_\mathrm{MT}\,{}\dot{M}_\mathrm{d})} &\mathrm{if}\,{}{\rm the}\,{} \mathrm{accretor}\,{} \mathrm{is}\,{} \mathrm{a}\,{} \mathrm{compact}\,{} \mathrm{object}, \\
  - f_\mathrm{MT}\,{} \dot{M}_\mathrm{d}&\mathrm{otherwise}, 
  \end{array}
  \right.
    \label{eq:maccrlo}
\end{equation}
 where $\dot{M}_{\rm Edd}$ is the Eddington rate (Eq.~67 of \citealt{hurley2002}), $\dot{M}_\mathrm{d}$ is the mass-loss rate of the donor star, and $f_{\rm MT}\in[0,1]$ is the mass accretion efficiency; here, we use $f_{\rm MT}=0.5$. 
Furthermore, we assume that the mass not accreted during the Roche-lobe overflow is lost from the vicinity of the accretor as an isotropic wind (isotropic re-emission). At the onset of Roche-lobe overflow, \sevn{} circularises the orbit at periastron.

To model a common-envelope phase, we assume an efficiency parameter $\alpha_{\rm CE}=1$ and estimate the envelope binding energy using the same $\lambda_\mathrm{CE}$ formalism as in \cite{claeys2014}. 

We model the final fate of intermediate-mass and high-mass stars as described in \cite{Iorio2023}. In particular, we use the rapid formalism by \cite{fryer2012} for core-collapse supernovae, we model electron-capture supernovae as in \cite{giacobbo2019}, and (pulsational) pair-instability supernovae according to \cite{mapelli2020}. Compact objects receive a natal kick at their birth. In our models,  we randomly draw the natal kick magnitude as \citep{giacobbo2020}:
\begin{equation}
V_{\rm kick}=f_{\rm H05}\,{}\frac{\langle{}M_{\rm NS}\rangle{}}{M_{\rm rem}}\,{}\frac{M_{\rm ej}}{\langle{}M_{\rm ej}\rangle},
\end{equation}    
where $\langle{}M_{\rm NS}\rangle{}$ and $\langle{}M_{\rm ej}\rangle$ are the average neutron star  mass and ejecta mass from single stellar evolution, respectively, while $M_{\rm rem}$ and $M_{\rm ej}$ are the compact object mass and the ejecta mass \citep{giacobbo2020}. The term $f_{\rm H05}$ is a random number drawn from a Maxwellian distribution with  one-dimensional root mean square $\sigma_\mathrm{kick}=265 \ \mathrm{km}\,{} \mathrm{s}^{-1}$, coming from a fit to the proper motions of 73 young pulsars ($<3$ Myr) in the Milky Way \citep{hobbs2005}.   In this formalism, stripped  and ultra-stripped  supernovae  result in lower kicks with respect to the other explosions, owing to the lower amount of ejected mass $M_{\rm ej}$ \citep{bray2016,bray2018}. 
BHs originating from a direct collapse receive zero natal kicks from this mechanism. 
We report the \sevn{} input parameter list in Zenodo at \url{https://doi.org/10.5281/zenodo.7736309} \citep{Costa2023_zenodo}.

\subsection{Tracks and single star evolution}
\begin{figure*}
	\includegraphics[width=\textwidth]{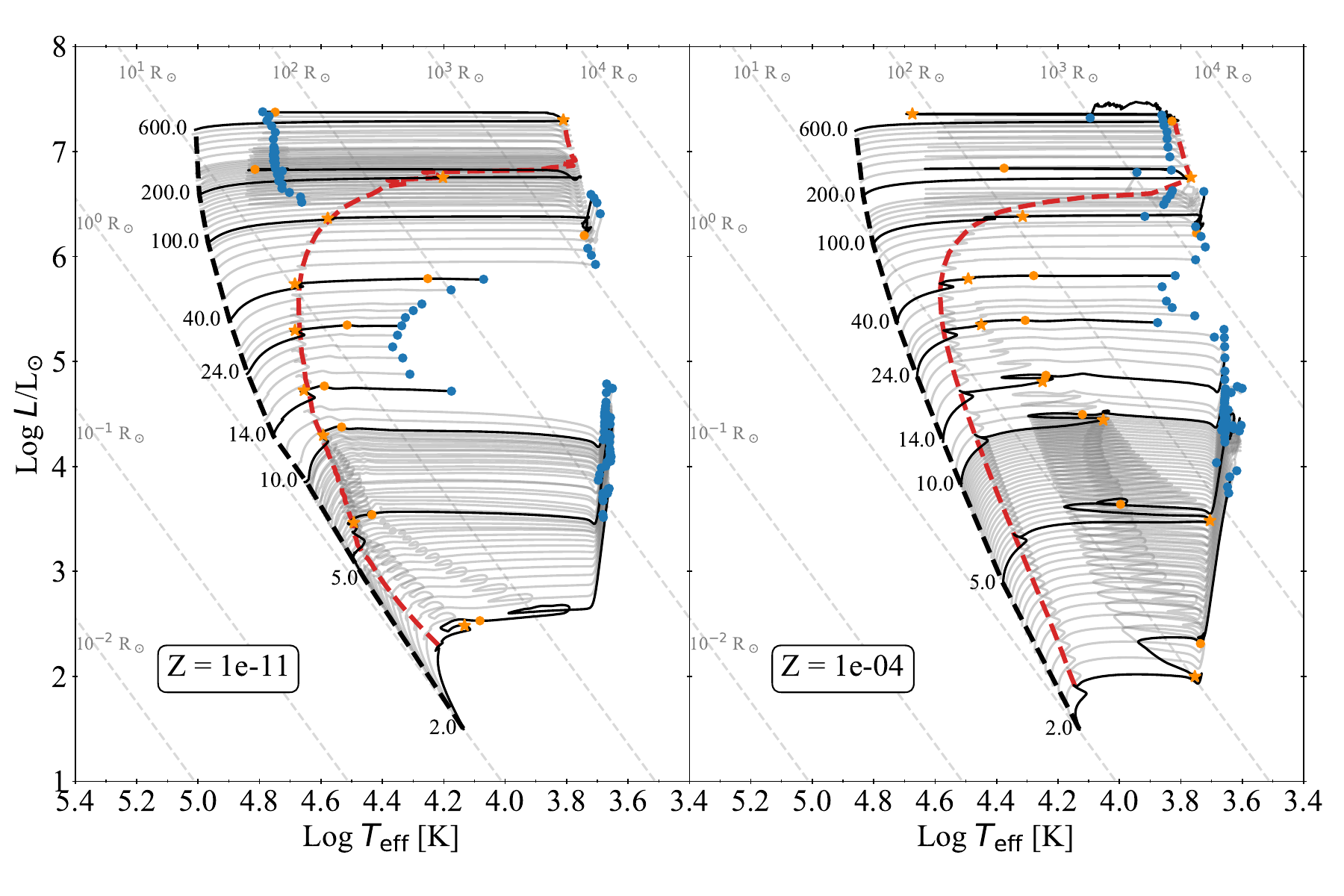}
    \caption{Hertzsprung–Russell (HR) diagram of Pop.~III (left) and Pop.~II (right) tracks. The black thick lines show the evolution of selected tracks with \Mz{}{}= 2, 5, 10, 14, 24, 40, 100, 200, and 600 M$_\odot$. All the other tracks are shown with solid grey lines. 
    The dashed black line indicates the \ac{ZAMS}. The red dashed line indicates the end of the \ac{MS}. The orange stars (circles) mark the beginning (end) of core He burning. The blue circles indicate the final position of the star in the diagram before the supernova. Diagonal grey dashed lines  correspond to  constant stellar radii in \Rsun.}
    \label{fig:HR}
\end{figure*}

\begin{figure*}
	\includegraphics[width=\textwidth]{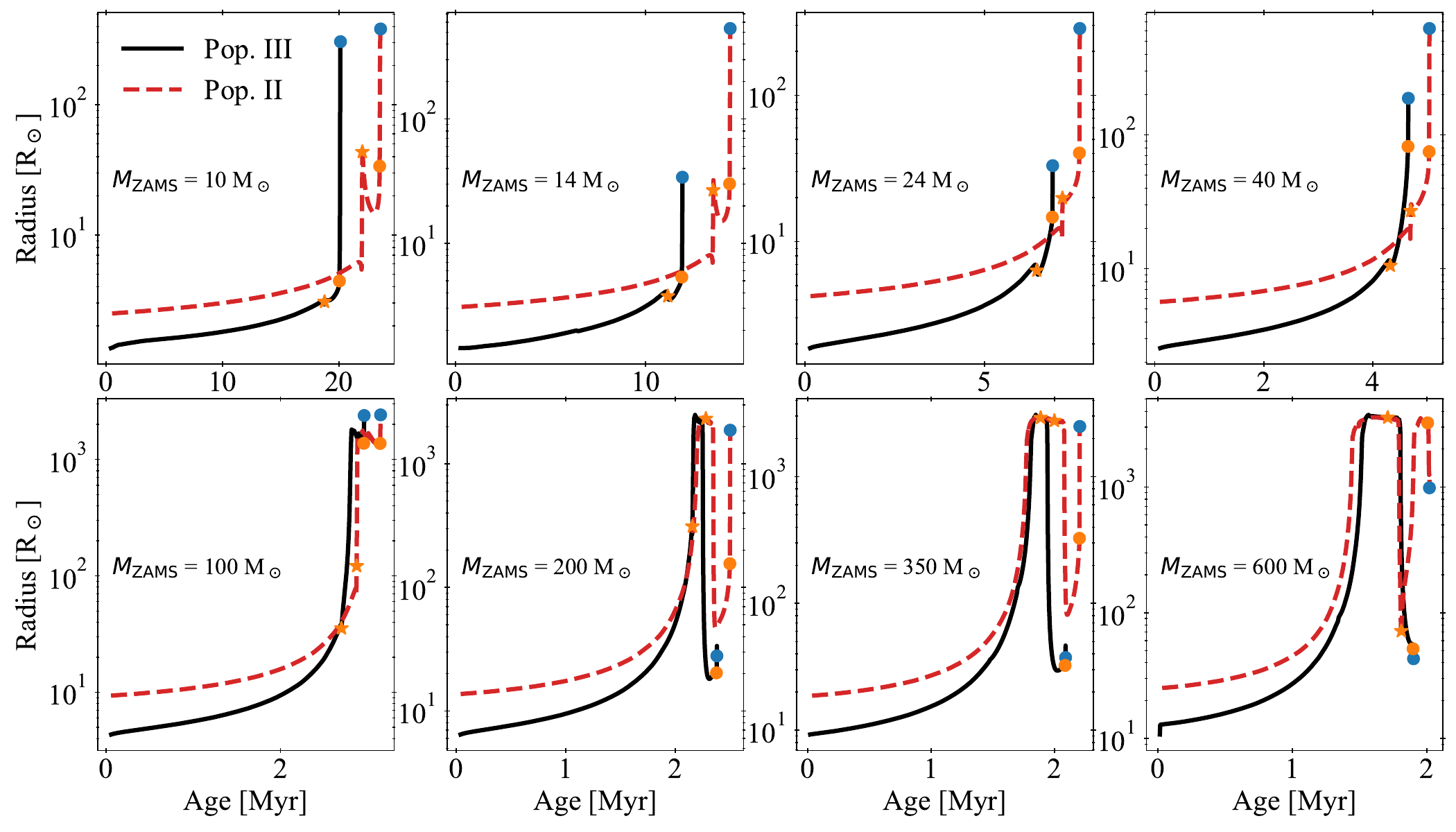}
    \caption{Radius versus age of some selected massive stars ($\Mz{}{} \geq 10~\Msun$). The solid black (dashed red) lines indicate Pop. III (Pop. II) stars. The orange stars (circles) mark the beginning (end) of core He burning. The blue circles indicate the final radius of the star before the supernova.}
    \label{fig:RvsAge}
\end{figure*}
We generated sets of Pop.~III and Pop.~II  stellar tracks with the {\sc parsec} code \citep{Bressan2012, costa2021, Costa2022, nguyen2022}. 
Pop.~III stars have typical behaviours, which Pop.~II stars do not show in their evolution \citep{Cassisi1993, Marigo2001, Murphy2021}.
For instance, during the early \ac{MS}, Pop.~III stars cannot ignite the CNO tri-cycle because of the initial lack of carbon, nitrogen and oxygen. 
Pop.~III stars need very high central temperatures to reach pressure support just with the energy provided by the proton-proton (pp) chain. 
Depending on the stellar mass, the central temperature becomes so high that some carbon could be synthesized from the triple-$\alpha$ reaction (i.e.,  helium burning), even in the  \ac{MS}.
This leads the CNO tri-cycle to ignite and suddenly replace the pp chain as the main source of energy of the star \citep{Marigo2001, Murphy2021}. 
Moreover, due to the high central temperatures reached at the end of the \ac{MS}, Pop.~III stars have a smoother transition to the \ac{CHeB} phase with respect to more metal-rich stars.  These characteristic features of Pop.~III stars evolution arise at metallicity $Z\lesssim 10^{-10}$ \citep[][]{Cassisi1993}. Hence, here we assume a metallicity $Z = 10^{-11}$ for Pop.~III stars \citep[see also][]{tanikawa2021}. For Pop.~II stars, we take a metallicity $Z = 10^{-4}$.
 
We adopt the \citet{Caffau2011} solar partition of chemical elements. Each set has an initial mass at the zero-age main sequence (ZAMS), \Mz{}{}, which ranges from 2 to 600~\Msun. 
All tracks evolve until advanced evolutionary phases. Stars with $2 \leq M_{\rm ZAMS}/\Msun{}< 10$ reach the early-\ac{AGB} phase (post-core helium-burning phase characterized by the burning of helium in a shell above the CO core). Stars with $M_{\rm ZAMS} \geq 10\,{}\Msun$ evolve until advanced phases of the core oxygen burning or the beginning of the pair-instability regime. 
The Pop.~III set of tracks extends the collection of {\sc parsec} tracks, already used in \sevn{} \citep[][]{Iorio2023}, and will soon be made available online\footnote{\url{http://stev.oapd.inaf.it/PARSEC}}. 

We also computed new tracks of pure-He stars to extend the database to lower metallicities. We use tracks of pure-He stars to model  naked-He stars formed  via stripping during mass transfer  at low metallicity \citep{kruckow2018,spera2019,mapelli2020,Iorio2023,agrawal2023}.  The metallicity adopted is  $Z = 10^{-6}$, and the masses range from $M_{\rm ZAMS,He}= 0.36$ to $350\,{}\Msun$.
This metallicity is similar to the metal content we find in He cores of Pop.~III stars after the \ac{MS} phase. For instance, depending on the initial mass, the carbon mass fraction in the He cores goes from $7 \times 10^{-7}$ to $2.6 \times 10^{-6}$ for 2~\Msun\ and 600~\Msun, respectively. Therefore, we do not expect to have completely metal-free pure-He stars.

All the new tracks are computed with the same physical set-up as described in \citet{costa2021, Costa2022} for stellar winds, nuclear reaction network, opacities, and equation of state. 
In Pop.~III tracks, stellar winds are naturally quenched due to the lack of metals (particularly iron) which we accounted for \citep[see][]{chen2015}.
Concerning the stellar convection, we adopt the Schwarzschild criterion \citep{Schwarzschild1958} for defining the unstable region and the mixing-length theory \citep{Bohm-Vitense1958} with a solar-calibrated value for the parameter $\alpha_\mathrm{MLT}=1.74$ \citep{Bressan2012}. Above the convective core, we adopt a penetrative overshooting with a characteristic parameter of $\lambda_\mathrm{ov} = 0.5$ in units of pressure scale height, computed with the ballistic approach \citep{Bressan1981}. In this framework, $\lambda_\mathrm{ov}$ is the mean free path of the unstable element across the border of the convective region, and its value corresponds to about $f_\mathrm{ov} = 0.025$ in the exponential decay overshooting formalism \citep[in the diffusive scheme,][]{Herwig1997}.
We also included undershooting at the bottom of the convective envelope, with a characteristic distance of $\Lambda_\mathrm{env} = 0.7$ in pressure scale heights.
The undershooting could play a role in the ending fate of massive stars, triggering dredge-up episodes which may stabilize the star against  pair instability \citep{costa2021, volpato2023}. 
In the interpolation methods used in \sevn{}  \citep[described in detail in][]{Iorio2023}, we cut the evolution just before the early-\ac{AGB} or the ignition of core C burning. 

Figure~\ref{fig:HR} shows the two sets of tracks used in this work, Pop.~III and II stars, in the Hertzsprung–Russell (HR) diagram. 
Pop.~III stars begin their life as metal-free objects and, in the \ac{ZAMS}, are more compact and hotter than their Pop.~II counterparts.

Both Pop. III and Pop. II stars evolve towards the red part of the diagram during the \ac{MS}. 
Figure~\ref{fig:HR} also shows a clear trend of the star position at the end of the \ac{MS} as a function of the initial stellar mass. 
Pop. III stars with an $\Mz{}{} \leq 200$~\Msun\ end the \ac{MS} as \ac{BSG} stars, while, stars with $\Mz{}{} > 200$~\Msun\ complete the \ac{MS} as \ac{YSG} or \ac{RSG} stars. This trend is similar in Pop. II stars, but with a lower transition mass (about 150~\Msun). Such a trend for Pop. III stars has also been found by other authors \citep[e.g.,][]{tanikawa2021}, but for higher initial masses ($\Mz{}{} > 600$~\Msun). 
Stars that become \ac{RSG} during the \ac{MS} develop large convective envelopes, differently from stars that remain \ac{BSG}, which have mostly radiative envelopes. 
The transition mass that separates the two evolutionary pathways  depends on the convection treatment adopted. 
This peculiar evolution of the most massive Pop.~III and Pop.~II stars can dramatically affect the evolution of a binary system since they become giant stars with very large radii during the \ac{MS}, in which there is still no well-defined transition between the core and the envelope. 
Therefore, binary interactions in such cases may lead to early mergers. 

The post-\ac{MS} evolution of Pop.~III stars shows  several features in the HR diagram, which depend on  \Mz{}.
Due to the high central temperatures during the \ac{MS}, all the tracks ignite helium as \ac{BSG} stars shortly after the end of the \ac{MS}.
Stars with a mass $\Mz{}{} \leq 40~\Msun$ end the \ac{CHeB} phase in the blue side of the HR diagram. 
After the \ac{CHeB} phase, stars with an initial mass between $2 \leq \Mz{}{}/\Msun \leq 10$ evolve to the \ac{AGB}. 
Stars above 10~\Msun\ evolve through all the advanced phases up to the oxygen burning, but they die with different final configurations.
Stars in the mass range $10 < \Mz{}{}/\Msun < 14$ move to the red part of the HR diagram and explode as \ac{RSG}.
Stars in the mass range $14 \leq \Mz{}{}/\Msun \leq 40$ die as \ac{BSG}.
Similar behaviour in this mass range has been found by other authors \citep[e.g.][]{Marigo2001, tanikawa2021}.
Stars in the mass range $40 < \Mz{}{}/\Msun \leq 100$ deplete helium in the red part of the diagram and finish their evolution as \acp{RSG}.
Finally, stars with $\Mz{}{} > 100$~\Msun\ ignite helium as \acp{RSG}, become \acp{BSG} during the \ac{CHeB}, and remain \acp{BSG} until their final fate.

Concerning Pop. II stars, intermediate-mass stars ($2 \leq \Mz{}{}/\Msun < 10$) 
do a blue loop during the \ac{CHeB}, and then, after helium depletion, move to the AGB phase.
Stars in the mass range $10 < \Mz{}{}/\Msun \leq 100$ burn helium as \acp{BSG} before moving to the red part of the diagram.
While stars with an initial mass $\Mz{}{} \geq 100~\Msun$ ignite helium as \acp{RSG}, burn it as \acp{BSG} (blue loop again), and later move back to the red.
All massive stars ($\Mz{}{} \geq 10~\Msun$) burn all the elements up to oxygen and finally explode as \acp{RSG}.

Figure~\ref{fig:RvsAge} shows the comparison between the radius evolution of Pop.~III and Pop.~II stars. 
Pop.~III stars with $\Mz{}{} \leq 100~\Msun$ evolve and reach the \ac{RSG} stage before Pop.~II stars. 
Pop.~III and Pop.~II stars with a mass of $\ge{}200$~\Msun\ become \acp{RSG} with comparable timescales. 
Figure~\ref{fig:RvsAge} shows that Pop.~III stars with mass $\Mz{}{} > 100~\Msun$ reach the pre-supernova stage as compact \acp{BSG} stars, whereas Pop.~II stars explode as very large \ac{RSG} stars. Pop.~III stars evolve to the final pre-supernova stage faster than Pop.~II stars.

\subsection{Binary initial conditions}
\label{sec:bin_init_cond}

In this Section, we describe the initial conditions used in this work for computing binary-population catalogues.

\subsubsection{Initial mass function (IMF)}
\label{sec:IMF}

There is still no consensus about the \ac{IMF} of Pop.~III stars, although several papers suggest that it should be rather top-heavy with respect to that of local stars \citep[e.g.,][]{Chiosi1998, abel2002,bromm2004,yoshida2006,bromm2013, glover2013, Goswami2022}. The transition between a top-heavy and a bottom-heavy mass function likely happened in the metallicity range of Pop.~II stars \citep{chon2021,sharda2022}. Here, given the uncertainties, we consider the same set of possible IMFs for both Pop.~III and Pop.~II stars, as follows.

\begin{figure}
	\includegraphics[width=\columnwidth]{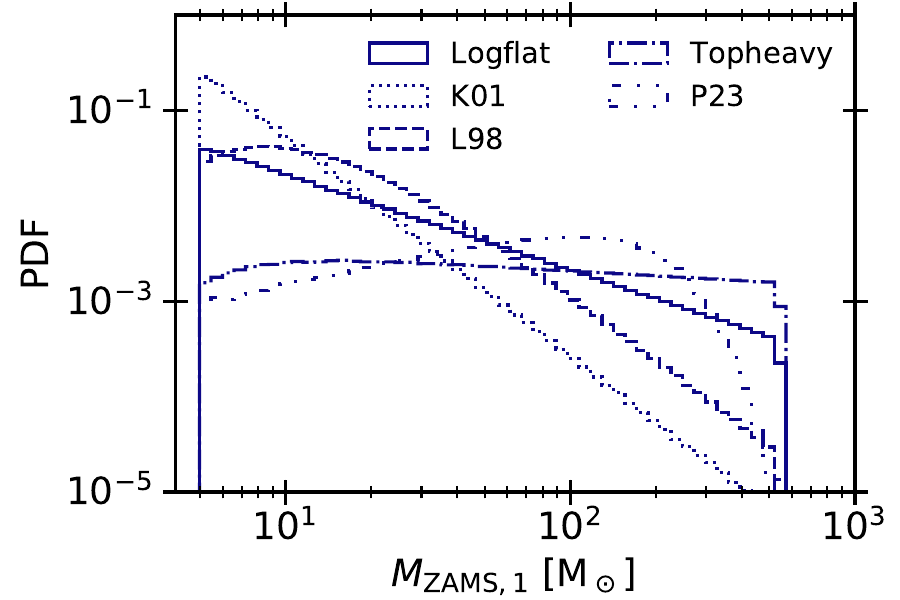}
    \caption{
    Initial mass distribution of the primary star ($M_{\rm ZAMS,1}$) in our models, as described in Section~\ref{sec:IMF} and Table~\ref{tab:IC}.}
    \label{fig:IC_IMF}
\end{figure}

\begin{figure*}
	\includegraphics[width=\textwidth]{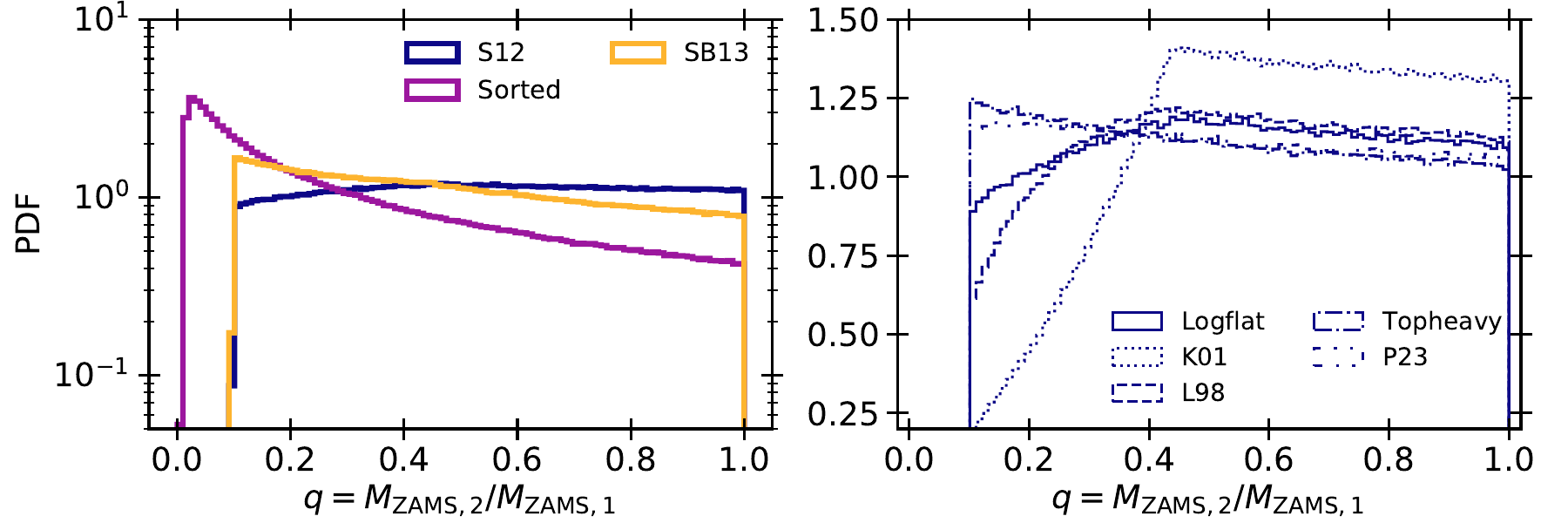}
    \caption{The left-hand panel shows our three initial mass-ratio ($q=M_{\rm ZAMS,2}/M_{\rm ZAMS,1}$) distributions (\citetalias{sana2012}, Sorted, and \citetalias{stacy2013}), calculated assuming a flat-in-log IMF for the primary mass.
    The right-hand panel shows the behaviour of the \citetalias{sana2012} mass-ratio distribution depending on the primary mass  function (flat-in-log, \citetalias{kroupa2001}, \citetalias{larson1998}, top-heavy, and \citetalias{Park2023}).
    The line style is the same as Fig. \ref{fig:IC_IMF}. See Table~\ref{tab:IC} for more details.}
    \label{fig:IC_q}
\end{figure*}

\begin{figure*}
	\includegraphics[width=\textwidth]{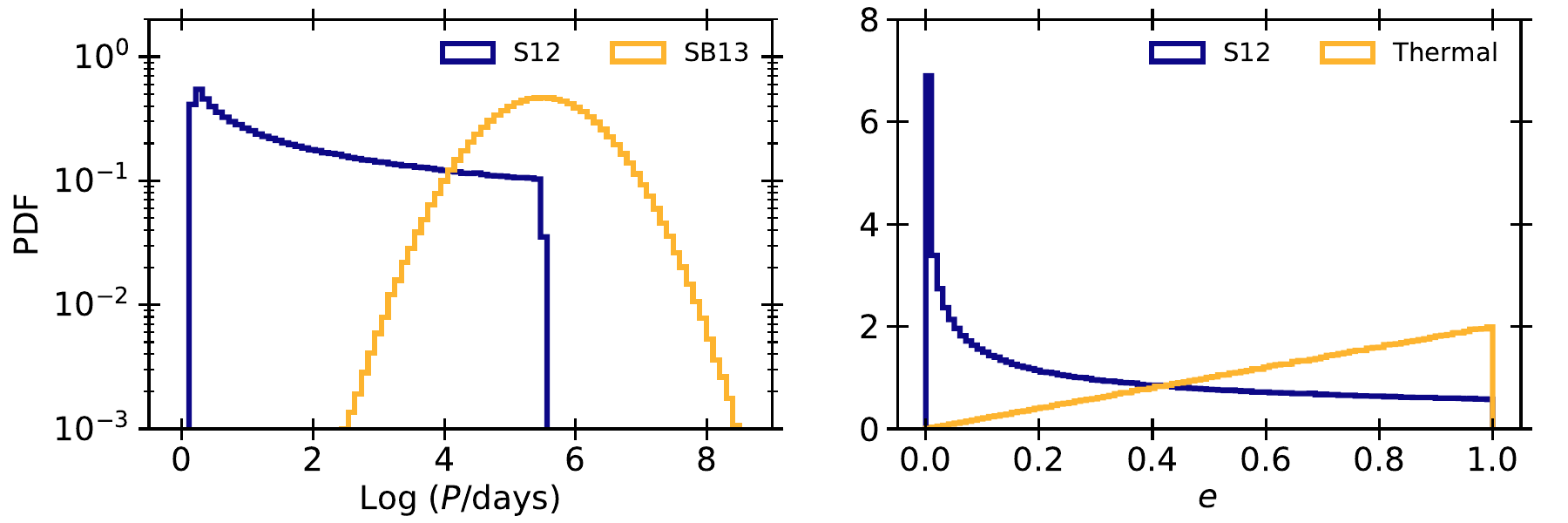}
    \caption{The left-hand and right-hand panels show the orbital period $P$ and eccentricity $e$ distributions adopted in our initial conditions (Table~\ref{tab:IC}).
    }
    \label{fig:IC_P_e}
\end{figure*}

\begin{itemize}
\item A \textit{flat-in-log} distribution \citep{stacy2013,susa2014,hirano2014,hirano2015,wollenberg2020,chon2021,tanikawa2021,jaura2022,prole2022}:
    \begin{equation}
        \xi(M_{\rm ZAMS}) \propto M_{\rm ZAMS}^{-1}.
    	\label{eq:IMF_Logflat}
    \end{equation}
    This IMF will be our fiducial model for Pop.~III stars. 
    \item A \citet{kroupa2001} \ac{IMF} (hereafter, \citetalias{kroupa2001}):
    \begin{equation}
        \xi(M_{\rm ZAMS}) \propto M_{\rm ZAMS}^{-2.3}.
    	\label{eq:IMF_Kroupa}
    \end{equation}
    This mass function is often adopted for stars in the low-redshift Universe and will be our fiducial model for Pop.~II stars.  
    With respect to the original \citetalias{kroupa2001}, which has a flatter slope for $M_{\rm ZAMS}<0.5$ M$_\odot$, here we assume a single slope because we do not generate ZAMS masses $<5$ M$_\odot$ from this distribution. 
    \item A \citet{larson1998} distribution (hereafter, \citetalias{larson1998}):
    \begin{equation}
        \xi(M_{\rm ZAMS}) \propto M_{\rm ZAMS}^{-2.35} \,{}e^{- M_\mathrm{cut1}/M_{\rm ZAMS}},
    	\label{eq:IMF_Larson}
    \end{equation}
    where $M_{\rm cut1} = 20$ M$_\odot$ \citep{valiante2016}.
    
    \item A \textit{top-heavy} distribution \citep[][]{stacy2013,jaacks2019,Liu2020}:
    \begin{equation}
        \xi(M_{\rm ZAMS}) \propto M_{\rm ZAMS}^{-0.17} \,{}e^{- M_\mathrm{cut2}^2/M_{\rm ZAMS}^2},
    	\label{eq:IMF_Topheavy}
    \end{equation}
    where $M_\mathrm{cut2} = 20$~M$_\odot$.

\item The distribution derived by  \citet[][hereafter, \citetalias{Park2023}]{Park2023}, based on hydro-dynamical simulations of Pop. III star formation, including radiative feedback from proto-stars and a diffuse weak X-ray background \citep[see also][]{Park2021a, Park2021b}:
    \begin{equation}
        \xi(M_{\rm ZAMS}) \propto 
        M_{\rm ZAMS}^{0.62} e^{- M_{\rm ZAMS}^2/M_\mathrm{cut3}^2},
    	\label{eq:IMF_Park}
    \end{equation}
    where $M_\mathrm{cut3} = 188$~M$_\odot$.
\end{itemize}
The IMF distributions adopted in this work are shown in Fig. \ref{fig:IC_IMF}.

\subsubsection{Mass ratio and secondary mass}
\label{sec:q}

We draw the ZAMS mass of the secondary star ($M_{\rm ZAMS,\,{}2}$)  according to three different distributions. 

\begin{itemize}
    \item We use the distribution of the mass ratio ($q=M_{\rm ZAMS,2}/M_{\rm ZAMS,1}$) from  \citet[][hereafter  \citetalias{sana2012}]{sana2012}:
    \begin{equation}
        \xi(q) \propto q^{-0.1} \; {\rm with} \; q \in [0.1, 1] \; {\rm and} \; M_{\rm ZAMS,2} \geq 2.2~{\rm M}_{\odot}.
	    \label{eq:Sana}
    \end{equation}
     This distribution is a fit to the mass ratio of O- and B-type binary stars in the local Universe \citep{sana2012}.   
    
    \item In the \textit{sorted}  distribution, we draw the ZAMS mass of the entire star population from the same \ac{IMF}, and then we randomly pair two stars from this distribution, imposing that $M_{\rm ZAMS,2}\leq{}M_{\rm ZAMS,1}$. 
    In this model,  the  minimum mass of the secondary is equal to that of  the primary (5~M$_\odot$) by construction. 
    
    \item The mass ratio distribution by \citet[][hereafter \citetalias{stacy2013}]{stacy2013}: 
    \begin{equation}
        \xi(q) \propto{} q^{-0.55}\; {\rm with} \; q \in [0.1, 1] \; {\rm and} \; M_{\rm ZAMS,2} \geq 2.2~{\rm M}_\odot.
    	\label{eq:q_stacy}
    \end{equation}
    This distribution was obtained from a fit to Pop.~III stars formed in 
    cosmological simulations (\citetalias{stacy2013}). 
\end{itemize}
The final mass ratio distribution also depends on the mass distribution of the primary star, as shown in Fig.~\ref{fig:IC_q}. 

\subsubsection{Orbital period}
\label{sec:P}

We consider two different distributions for the initial orbital period ($P$), as shown in the left-hand panel of Fig.~\ref{fig:IC_P_e}:
\begin{itemize}
    \item The distribution derived by \citetalias{sana2012} for O- and B- stars in the local Universe:  
    \begin{equation}
        \xi(\Pi) \propto{} \Pi^{-0.55} \quad {\rm with} \quad \Pi = \log (P/{\rm day}) \in [0.15, 5.5].
    	\label{eq:P_sana}
    \end{equation}
    
    \item The period distribution  found by \citetalias{stacy2013}:
    \begin{equation}
        \xi(\Pi) \propto{} \exp{\left[-(\Pi-\mu)^2/(2\,{}\sigma{}^2)\right]}.
    	\label{eq:P_stacy}
    \end{equation}
    This is a Gaussian distribution with $\mu{}=5.5$, and  $\sigma{}=0.85$, favouring long periods with respect to the \citetalias{sana2012} distribution.  While this distribution is likely affected by numerical resolution, which reduces the number of systems with short orbital periods, we decide to consider it as a robust upper limit to the orbital period of Pop.~III and II binary stars \citep[see also][]{Sugimura2020, Park2021b, Park2023}.
\end{itemize}

\subsubsection{Eccentricity}
\label{sec:e}

We compare two distributions for the orbital eccentricity, as shown in the right-hand panel of Fig.~\ref{fig:IC_P_e}:
\begin{itemize}
    \item the distribution obtained by  \citetalias{sana2012} and based on a sample of O- and B-type stars in the local Universe:
    \begin{equation}
        \xi(e) \propto e^{-0.42} \quad {\rm with} \; e \in [0, 1).
    	\label{eq:ecc_sana}
    \end{equation}

    \item The thermal distribution, adopted  for Pop.~III binaries by, e.g., \cite{Kinugawa2014,Hartwig2016,tanikawa2021}:
    \begin{equation}
        \xi(e) =2\,{} e\quad {\rm with} \; e\in [0, 1).
    	\label{eq:ecc_therm}
    \end{equation}
    This eccentricity distribution favours highly eccentric systems, at variance with Eq.~\ref{eq:ecc_sana}. 
    Recent hydro-dynamical simulations \citep{Park2021b, Park2023} suggest that Pop. III binary stars form preferentially with high orbital  eccentricity, favouring the distribution in Eq.~\ref{eq:ecc_therm} with respect to Eq.~\ref{eq:ecc_sana}.
\end{itemize}

\subsubsection{Input catalogues}
\label{sec:ICc}

\begin{table*}
    \caption{Initial conditions.}  
        \begin{threeparttable}
        \begin{tabular}{cccccccc}
            \hline
            Model &  $M_{\rm ZAMS,1}$  & $M_{\rm ZAMS}$   & Mass ratio $q$     & Period $P$      & Eccentricity $e$ &  $N$ [$\times{}10^7$] & Total mass [$\times{}10^9$~\Msun] \\ 
            \hline
            LOG1     & Flat in log  & -- & \citetalias{sana2012}           & \citetalias{sana2012}         & \citetalias{sana2012}     & 1.45 & 2.59       \\
            LOG2     & Flat in log  & --& \citetalias{sana2012}           & \citetalias{stacy2013}           & Thermal  & 1.45 &  2.58    \\
            LOG3     & --  & Flat in log & Sorted         & \citetalias{sana2012}           & \citetalias{sana2012}     & 1.38 &  3.19    \\
             LOG4   & Flat in log  & -- & \citetalias{stacy2013}           & \citetalias{sana2012}           &   Thermal  & 1.53 & 2.60      \\
            LOG5     & Flat in log & -- & \citetalias{stacy2013}           & \citetalias{stacy2013}           &  Thermal  & 1.53 & 2.60 \\
            \hline
            KRO1     & \citetalias{kroupa2001}  & -- & \citetalias{sana2012}           & \citetalias{sana2012}           & \citetalias{sana2012}     & 5.23 (2.00$\dagger$)  & 1.35 (0.89$\dagger$)      \\
            KRO5     & \citetalias{kroupa2001}   & -- & \citetalias{stacy2013}           & \citetalias{stacy2013}           & Thermal  & 6.11 (2.00$\dagger$)   & 1.52 (0.93$\dagger$)  \\ 
            \hline
            LAR1     & \citetalias{larson1998}  & -- & \citetalias{sana2012}           & \citetalias{sana2012}            & \citetalias{sana2012}     & 2.00    & 1.20     \\
            LAR5     & \citetalias{larson1998}  & -- &  \citetalias{stacy2013}           & \citetalias{stacy2013}            & Thermal  &  2.27 (2.00$\dagger$)  & 1.30 (1.24$\dagger$) \\ 
            \hline
            TOP1     & Top heavy & -- & \citetalias{sana2012}          & \citetalias{sana2012}           & \citetalias{sana2012}      & 1.05 &  4.16     \\
            TOP5     & Top heavy & -- & \citetalias{stacy2013}          & \citetalias{stacy2013}           & Thermal   &  1.07 & 4.03  \\ 
            \hline
            PAR1     & \citetalias{Park2023} & -- & \citetalias{sana2012}          & \citetalias{sana2012}           & \citetalias{sana2012}     & 1.05 &  2.35     \\
            PAR5     & \citetalias{Park2023} & -- & \citetalias{stacy2013}          & \citetalias{stacy2013}           & Thermal   &  1.06 & 2.28  \\ 
            \hline
        \end{tabular}
        \footnotesize
            Column~1 reports the model name. Columns 2 describes how we generate the ZAMS mass of the primary star (i.e., the most massive of the two members of the binary system). Column~3 describes how we generate the ZAMS mass of the overall stellar population (without differentiating between primary and secondary stars). We follow this procedure only for model LOG3 (see the text for details). Columns~3, 4, and 5 specify the distributions we used to generate the mass ratios, the orbital periods and the orbital eccentricity. See Section~\ref{sec:bin_init_cond} for a detailed description of such distributions. The last two columns report the total number and the total mass of the of simulated binaries, respectively. $\dagger$The ICs for such models are under-sampled, the actual number of simulated systems and their total mass is reported in parentheses. See main text for additional details.
    \end{threeparttable}
    \label{tab:IC} 
\end{table*}

We build 13 different input catalogues by varying the aforementioned distributions  of
the \ac{IMF}, $q$, $P$, and $e$. 
For each of these catalogues, we consider the two metallicities for Pop.~III and Pop.~II, i.e. $Z = 10^{-11}$ and $ 10^{-4}$, respectively.

We set the total number of generated binaries to  obtain   $10^7$ binaries in the high-mass regime ($M_{\rm ZAMS,2} \geq 10~{\rm M}_\odot$, and $M_{\rm ZAMS,1} \geq 10~{\rm M}_\odot$ by construction). 
For the models that draw the primary mass from \citetalias{kroupa2001} and \citetalias{larson1998}  (Section~\ref{sec:IMF}), we limit the total number of generated binaries to $2\times{}10^7$ (consisting of $10^7$ binaries in the high- and low-mass range, respectively). 
As a consequence, the low-mass regime ($M_{\rm ZAMS,2} \leq 10~{\rm M}_\odot$) is under-sampled by a factor of $\approx4-5$  for \citetalias{kroupa2001} and $\lessapprox1.2$ for \citetalias{larson1998}. We take into account the incomplete sampling of the initial conditions by performing  an a posteriori over-sampling 
of the simulated binaries with $M_{\rm ZAMS,2} \leq 10~{\rm M}_\odot$.
This ensures a good sampling of the high-mass regime and reduces  stochastic fluctuations  \citep[e.g.,][]{Iorio2023}. 

Table~\ref{tab:IC} lists the properties of our input catalogues. 
We label our input catalogues by taking the IMF name and adding a number that indicates the distribution of mass ratios, periods,  and eccentricities. Therefore, the LOG, KRO, LAR, TOP,  and PAR catalogues adopt the flat-in-log, \citetalias{kroupa2001}, \citetalias{larson1998}, top-heavy,  and \citetalias{Park2023} IMF, respectively.

In all our models but LOG3 (Table~\ref{tab:IC}), we randomly sample the ZAMS mass of the primary star $M_{\rm ZAMS,1}$ (i.e., the ZAMS mass of the most massive member) of the binary system  in the range  $[5,\,{}550]$~M$_\odot$, according to one of the five aforementioned distributions.
We then randomly sample the ZAMS mass of the secondary star ($M_{\rm ZAMS,2}$) based on the mass ratio distributions described in Section~\ref{sec:q}. We assume that the secondary mass can be as low as $2.2$~M$_\odot$.

In model LOG3, we instead randomly sample the entire \ac{IMF} in the range  $M_{\rm ZAMS}\in{[5,\,{}550]}$~M$_\odot$, according to the LOG distribution. We then randomly pair the generated stellar masses. The primary (secondary) star is thus the component with the higher (lower) initial mass (see model sorted in Section~\ref{sec:q}).
Hereafter, we assume the models LOG1 and KRO1 as our fiducial case for Pop.~III and Pop.~II stars, respectively.

\section{Results}
\label{sec:results}

\subsection{Black holes from single star evolution}
\label{subsec:SSE}

\begin{figure}
\includegraphics[width=\columnwidth]{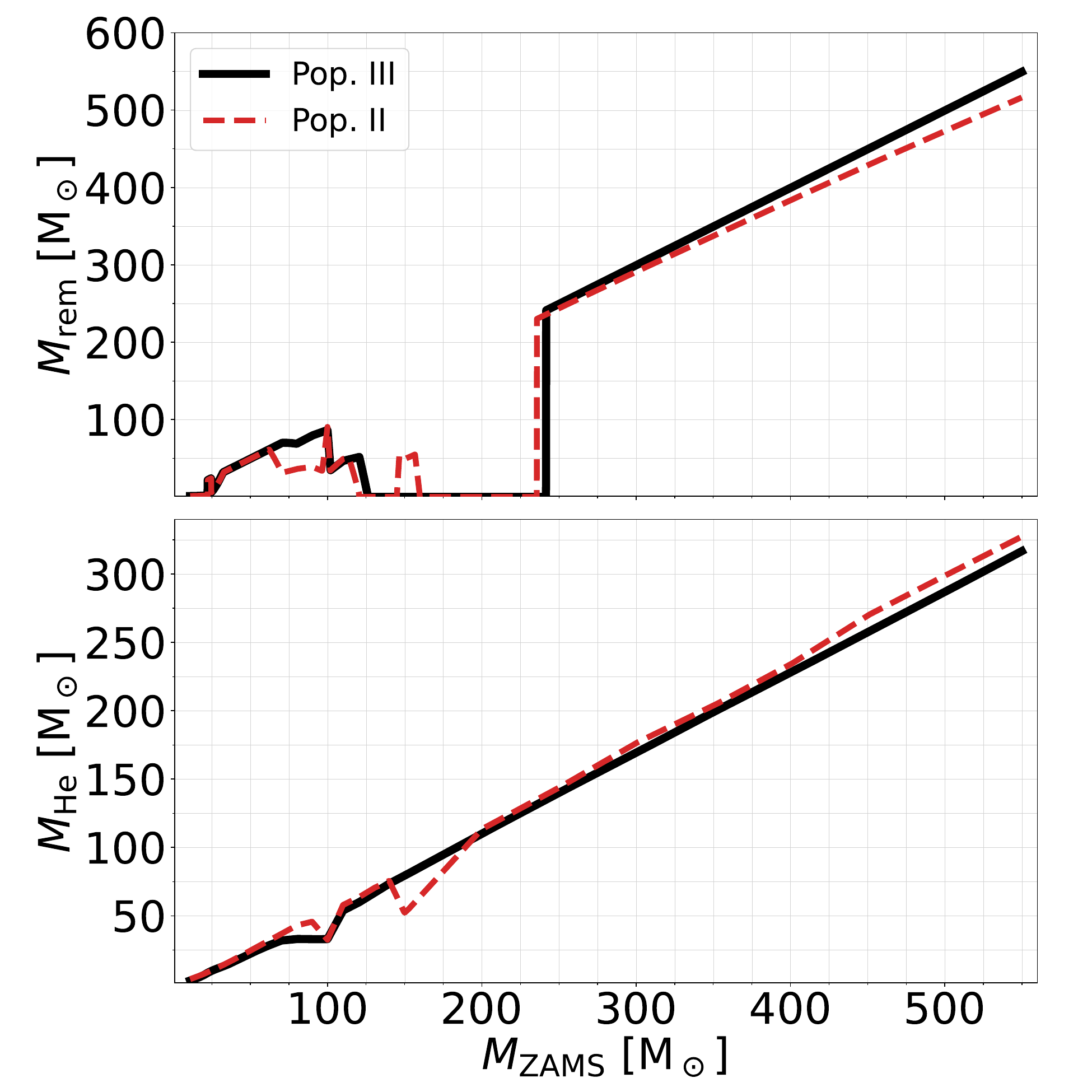}
\caption{The upper and lower panel show the mass of the compact remnant, and the mass of the He core at the onset of core collapse  as a function of the initial mass, \Mz{}{}.
    Black solid and red dashed lines refer to Pop.~III and Pop.~II stars, respectively.}
    \label{fig:MassSpec}
\end{figure}

Figure~\ref{fig:MassSpec} shows the mass of the compact remnant (\Mrem{}{}) as a function of the ZAMS mass (\Mz{}{}) for Pop.~II and Pop.~III stars evolved via single stellar evolution. Pop.~II and ~III stars evolving via single star evolution produce similar He core masses and, thus, similar compact remnant masses. The only differences are (i) in the range between pulsational pair instability and pair instability ($M_{\rm ZAMS}\in[60,\,{}170]$~M$_\odot$), where envelope overshooting can cause dredge-up episodes, and (ii) at extremely high BH masses ($M_{\rm rem}>400$ M$_\odot$), where Pop.~II stars suffer from slightly higher mass loss rates.

In the region between pulsational pair instability and pair instability ($M_{\rm ZAMS}\in[60,\,{}170]$ M$_\odot$), the He-core mass does not grow monotonically, especially in the case of Pop.~II stars. 
The core decrease in some mass ranges is caused by dredge-up episodes triggered by envelope undershooting \citep{costa2021}. 
Different choices for the convection parameters, such as the core overshooting ($\lambda{}_{\rm ov}=0.5$ in our models), can change the behaviour and the occurrence of dredge-up episodes. For instance, Pop.~II stars (Z~=~0.0001) with $\lambda{}_{\rm ov}=0.4$ show a monotonic trend of the He core mass \citep[see discussion in ][]{Iorio2023}.
In the models presented in this work, for ZAMS mass $M_{\rm ZAMS}\in{}[145-160]$ M$_\odot$ we expect an "island" of massive BH formation for Pop.~II stars inside a region of pair instability. 
This happens because a dredge-up episode reduces the mass of the He and CO core below the threshold for pair instability in this range for Pop.~II stars, but not for Pop.~III stars. 

The maximum mass of a Pop.~III BH below the pair-instability mass gap is 86 M$_\odot$ for the adopted pair-instability model. Similarly, the maximum mass of a Pop.~II BH below the mass gap is 91 M$_\odot$. In both cases, this mass is reached for a  ZAMS mass $\approx{100-105}$ M$_\odot$. Below the mass gap, 
our models predict several sharp features in the BH mass spectrum because of dredge-up episodes that affect the He core mass in this range. The mass spectrum in this region is  maximally sensitive to several details of the input physics that are highly uncertain (e.g., core overshooting, nuclear reaction rates), as already discussed in previous papers \citep[e.g.,][]{leung2019,farmer2019,farmer2020,mapelli2020,costa2021,woosley2021,vink2021}.

In our models, the upper edge of the mass gap is at $M_{\rm ZAMS}\approx{242}$ M$_\odot$ and $\approx{236}$ M$_\odot$ for Pop.~III and II stars, respectively. Above the mass gap,  both Pop.~III and ~II  stars produce intermediate-mass BHs from direct collapse. The mass of a BH born from a Pop.~III star in this regime is very similar to that of a BH formed by a  Pop.~II star with the same ZAMS mass, because stellar winds are already extremely quenched at $Z=10^{-4}$.

The maximum BH mass in our models is $M_{\rm rem} \approx 545$ M$_\odot$ ($\approx{510}$ M$_\odot$) for Pop.~III (Pop.~II) stars, corresponding to a ZAMS mass $M_{\rm ZAMS}=550$ M$_\odot$. 
We obtain these masses with the optimistic assumption that the residual H-rich envelope of the progenitor star collapses to a BH entirely when the star collapses. 
A fraction of the H-rich envelope mass might be lost even in the case of a failed explosion, because of shocks induced by the emission of neutrinos \citep[e.g.,][]{fernandez2018,Renzo2020b}. 

\subsection{Binary evolution}\label{sec:results_BSE}

\begin{figure}
\includegraphics[width=0.5\textwidth]{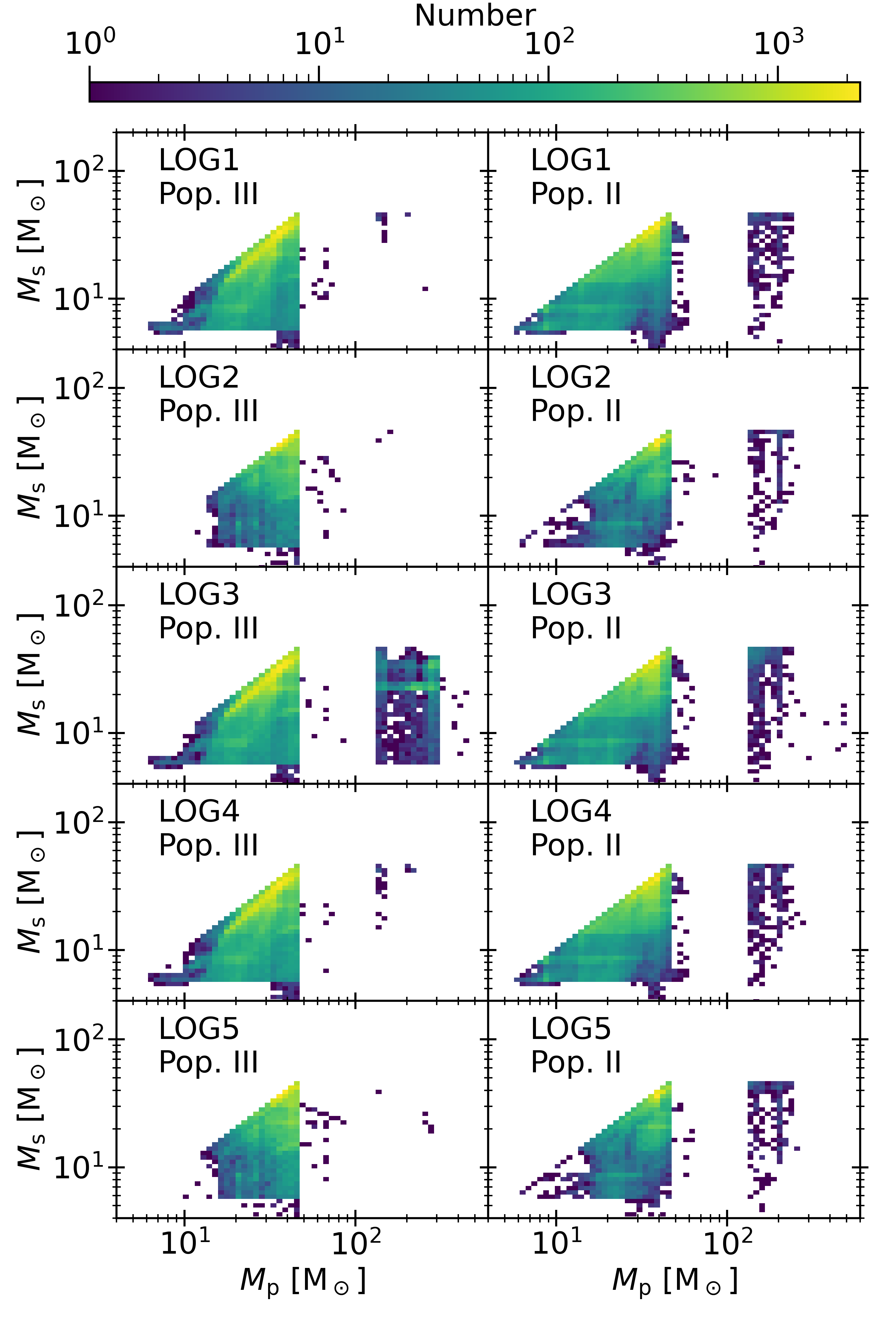}
    \caption{Distribution of secondary (\M{s}) versus primary (\M{p}) mass of all BBH mergers in our simulations \textsc{LOG1--5}. Left (Right): Pop.~III (II) stars. 
    The colour bar indicates the number of BBHs in each cell.}
    \label{fig:All_merging1}
\end{figure}

\begin{figure}
\includegraphics[width=0.5\textwidth]{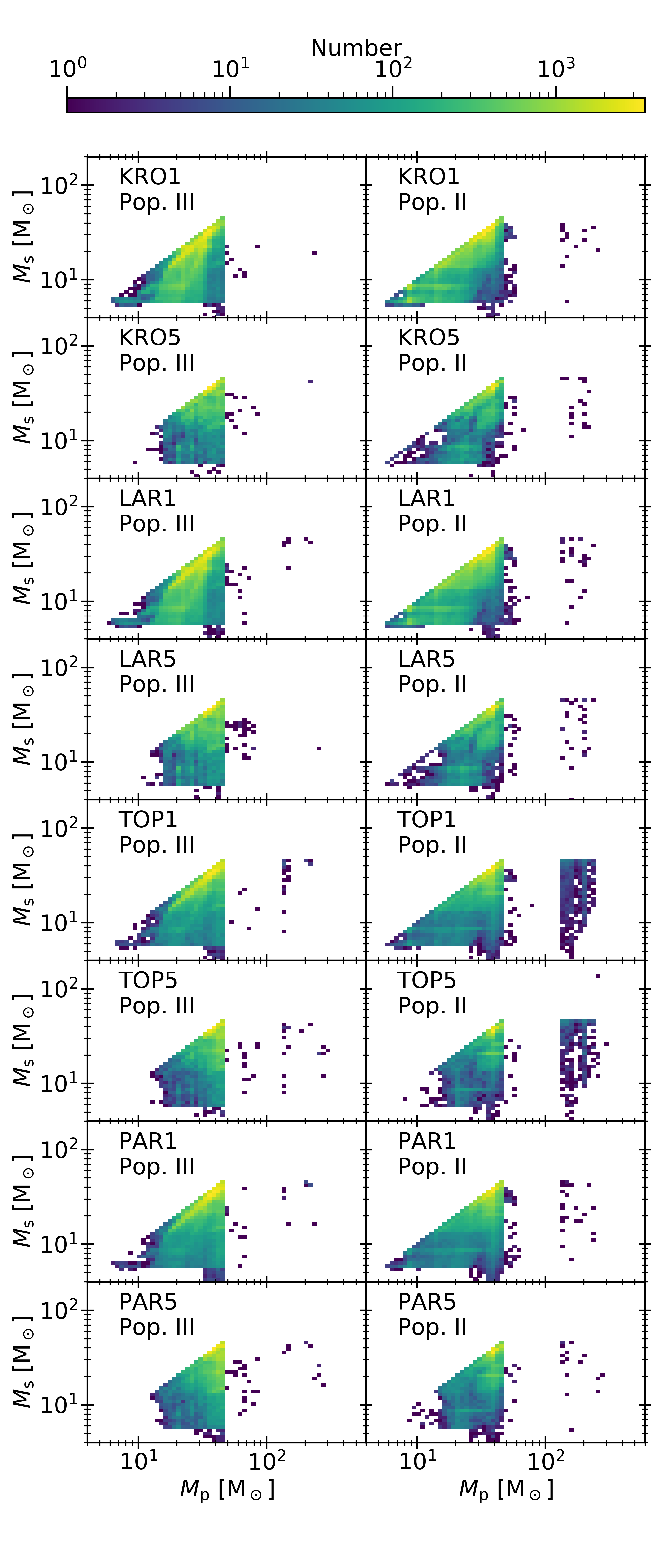} 
    \caption{
    Same as Fig.~\ref{fig:All_merging1} but for models KRO1, KRO5, LAR1, LAR5, TOP1, TOP5, PAR1, and PAR5}.
    \label{fig:All_merging2}
\end{figure}

Figures~\ref{fig:All_merging1} and \ref{fig:All_merging2} show the secondary BH mass (i.e. the mass of the least massive BH) versus the primary BH mass (i.e. the mass of the most massive BH) for all our simulated BBH mergers. The masses of BBHs born from Pop.~III stars are qualitatively similar to those of BBHs born from Pop.~II stars, for all the considered models. 

\begin{figure*}
\includegraphics[width=\textwidth]{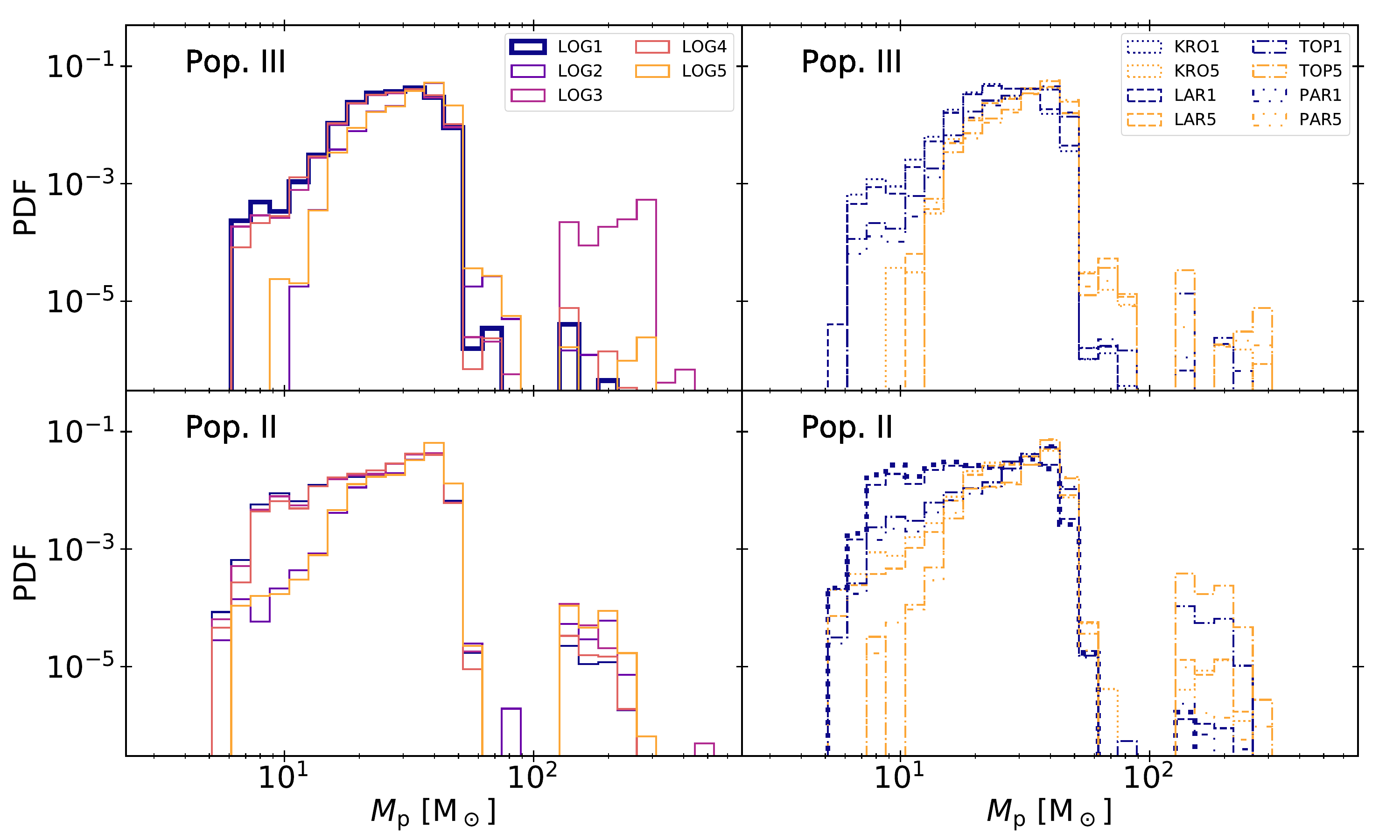}
    \caption{
    Distribution of the primary BH mass (i.e., the most massive BH in each binary system) in the simulated BBH mergers. Upper (lower) panel: Pop.~III (Pop.~II) stars. The left-hand panels show the models with the flat-in-log IMF, while the right-hand panels show all the other models. The fiducial models (i.e. LOG1 for Pop. III and KRO1 for Pop. II stars) are highlighted with a thicker line.}
    \label{fig:mp}
\end{figure*}

Mergers with at least one component above the pair-instability mass gap are not common, and mergers  inside the gap are even rarer. In our binary simulations,  it is even difficult to identify sharp edges for the pair instability mass gap, because of dredge-up episodes and mass transfer (Figure~\ref{fig:mp}). Assuming that the pair-instability mass gap spans from 85 to 230 M$_\odot$, we find that BBH mergers with primary BH masses above the gap are up to $3.3$\% (LOG3) and up to $0.09$\% (TOP5) for Pop.~III and Pop.~II stars, respectively. With the same definition, mergers with primary BH mass inside the gap are up to $1.9$\% (LOG3) and  up to $2.4$\% (TOP5)
for Pop.~III and Pop.~II stars, respectively.
In general, Pop.~II  stars seem to produce more BBH mergers with primary BH mass above the gap with respect to Pop.~III stars, with the exception of model LOG3 (Fig.~\ref{fig:mp}). Furthermore, no secondary BHs in our BBH mergers have mass  inside or above the gap. 

The most common primary BH masses are around 30--40~M$_\odot$ (Fig.~\ref{fig:mp}). There is a dearth of low-mass primary BHs ($8-10$ M$_\odot$) with respect to LVK mergers \citep{abbottO2popandrate,abbotO3apopandrate,abbottO3bpopandrate} in all of our runs, even KRO1. This is a consequence of the negligible mass loss rate and relatively compact stellar radii at such low metallicity. 

\begin{figure*}
\includegraphics[width=\textwidth]{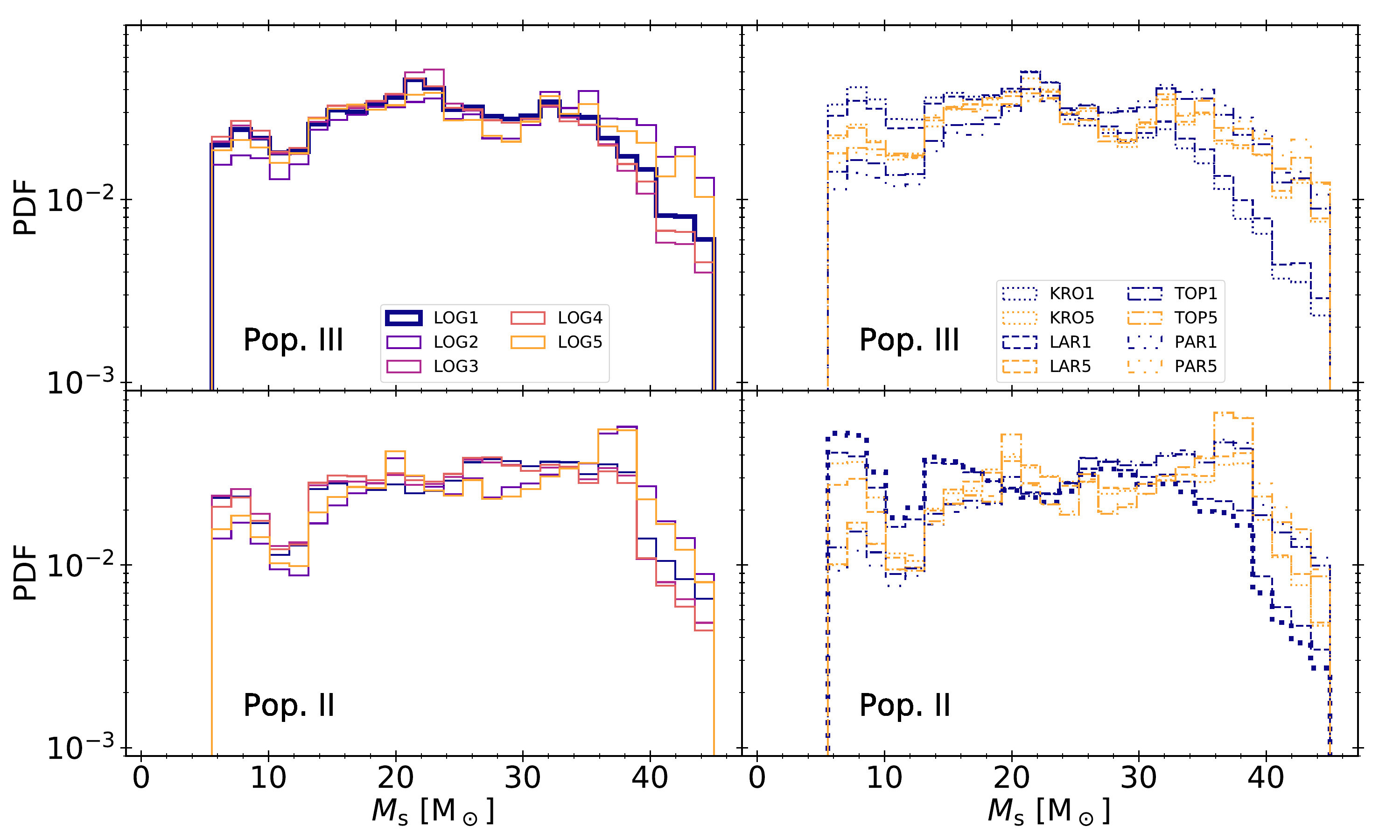} 
    \caption{Same as Fig.~\ref{fig:mp}, but for the secondary BH mass (i.e., the least massive BH of each binary system).}
    \label{fig:ms}
\end{figure*}

Our models show a  preference for equal mass systems but also a non-negligible contribution from unequal mass mergers (Figs.~\ref{fig:All_merging1} and ~\ref{fig:All_merging2}).  The secondary BH mass is always $M_{\rm s}\leq{}45$ M$_\odot$ for both Pop.~II and III stars. Figure~\ref{fig:ms} highlights  some differences between Pop.~II and III BBHs. For example, the most common secondary BH mass for Pop.~III stars is $\sim{20}$ M$_\odot$, while for Pop.~II stars it is either $\sim{8-10}$~M$_\odot$ (KRO1, LAR1), or $\sim{35-38}$~M$_\odot$ (LOG2, LOG5, TOP5), depending on the model.

\begin{figure*}
\includegraphics[width=\textwidth]{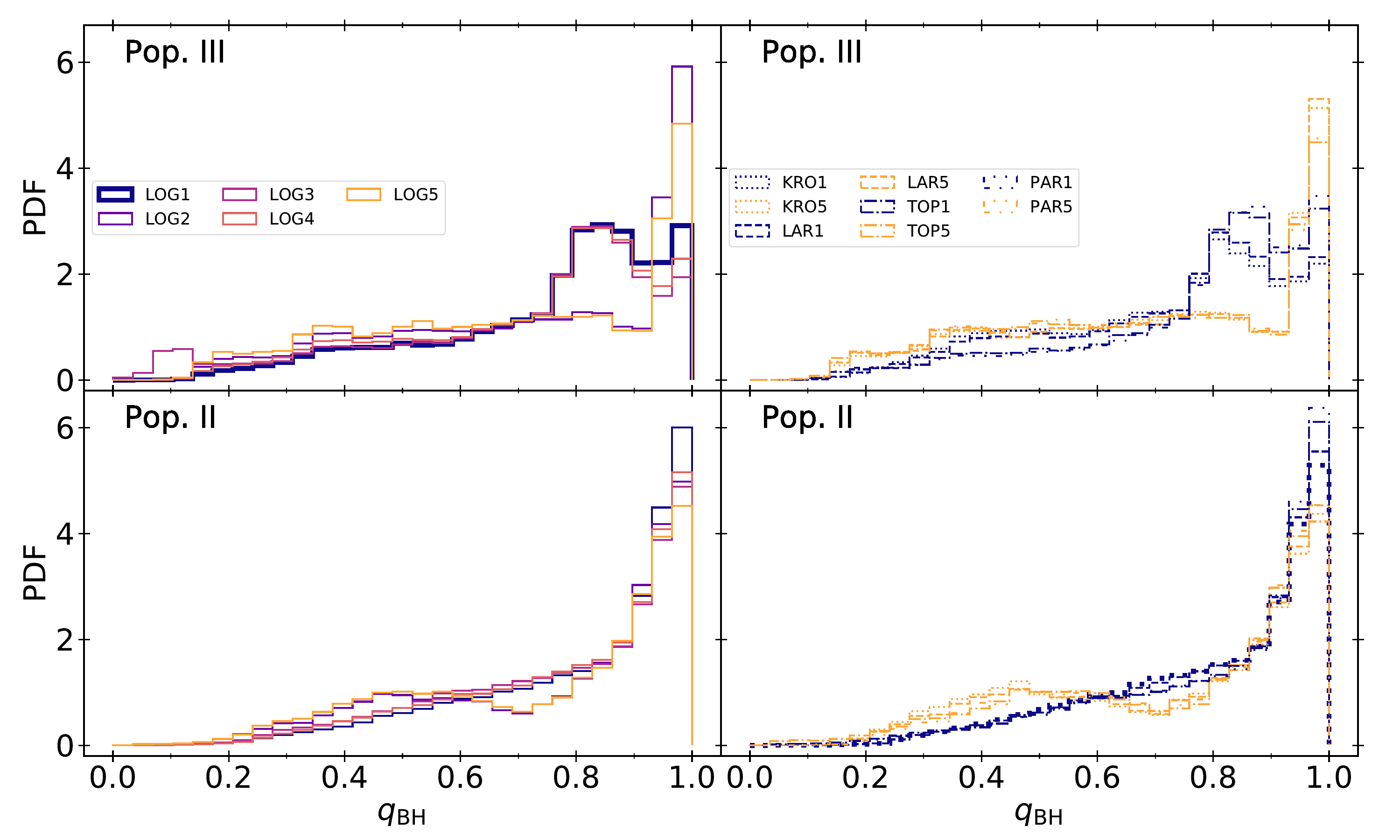} 
    \caption{Same as Fig.~\ref{fig:mp}, but for the mass ratio  $q_{\rm BH}=M_{\rm s}/M_{\rm p}$ between the secondary and primary BH.}
    \label{fig:q}
\end{figure*}

We find another interesting difference between Pop.~III and II BBHs if we look at the mass ratio (Fig.~\ref{fig:q}). In the case of Pop.~II stars, equal-mass BBHs are the most common systems regardless of the model, even if models LOG5, KRO5, LAR5, TOP5 and PAR5 show a mild secondary peak for $q\sim{0.4-0.5}$. In contrast, for Pop.~III stars, the most common BBH mass ratio is $\sim{0.8-0.9}$ for the models LOG1, LOG3, LOG4, KRO1, LAR1, TOP1 and PAR1, i.e. for all the models adopting the \citetalias{sana2012} initial period distribution. This is a consequence of  the dominant evolutionary channels in such models (see Section~\ref{sec:formation_ch}).

\begin{figure}
\includegraphics[width=0.8\columnwidth]{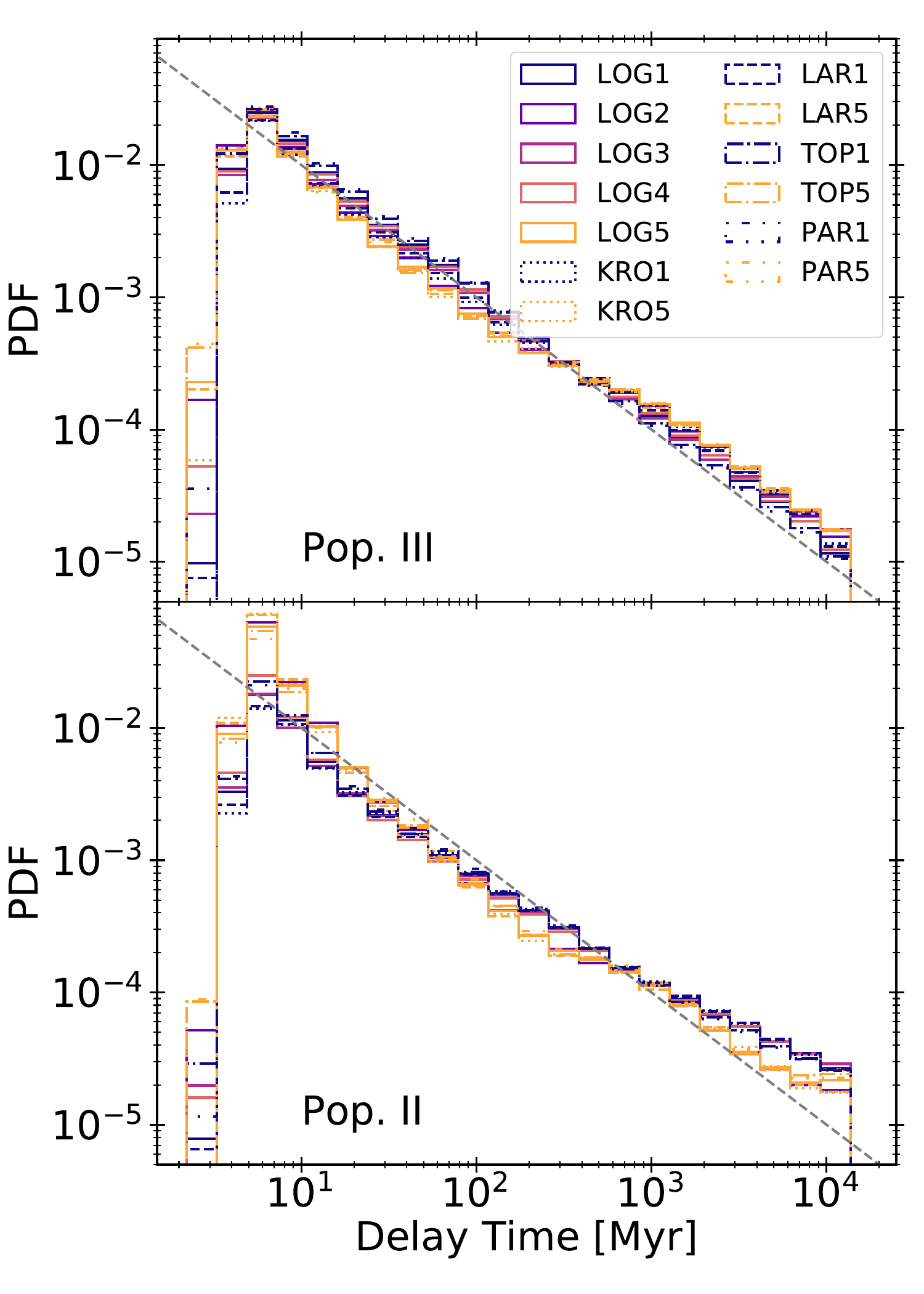} 
    \caption{
    Distribution of delay times, i.e. the time elapsed from the formation of the binary star to the merger of the two BHs. The gray dashed line shows the $\propto{}t^{-1}$ predicted trend \protect{\citep[e.g.,][]{dominik2012}}.}
    \label{fig:delay_time}
\end{figure}

Finally, the distribution of delay times $t_{\rm del}$ (i.e., the time elapsed between the formation of the binary system and the BBH merger) shows another difference between Pop.~III and II BBHs (Fig.~\ref{fig:delay_time}). All  Pop.~III models seem to match the trend $t_{\rm del}\propto{}t^{-1}$ between 3 and $10^4$ Myr. In contrast, some of the Pop.~II models (LOG2, LOG5, KRO5, LAR5, TOP5 and PAR5) show an excess of short delay times ($3-10$ Myr). The models showing this excess share the \citetalias{stacy2013} orbital period distribution. This feature is another signature of the formation channel, as we discuss in Section~\ref{sec:formation_ch}.

\section{Discussion}
\label{sec:disc}

\subsection{Formation channels of BBH mergers}\label{sec:formation_ch}

\begin{figure*}
\includegraphics[width=0.8\textwidth]{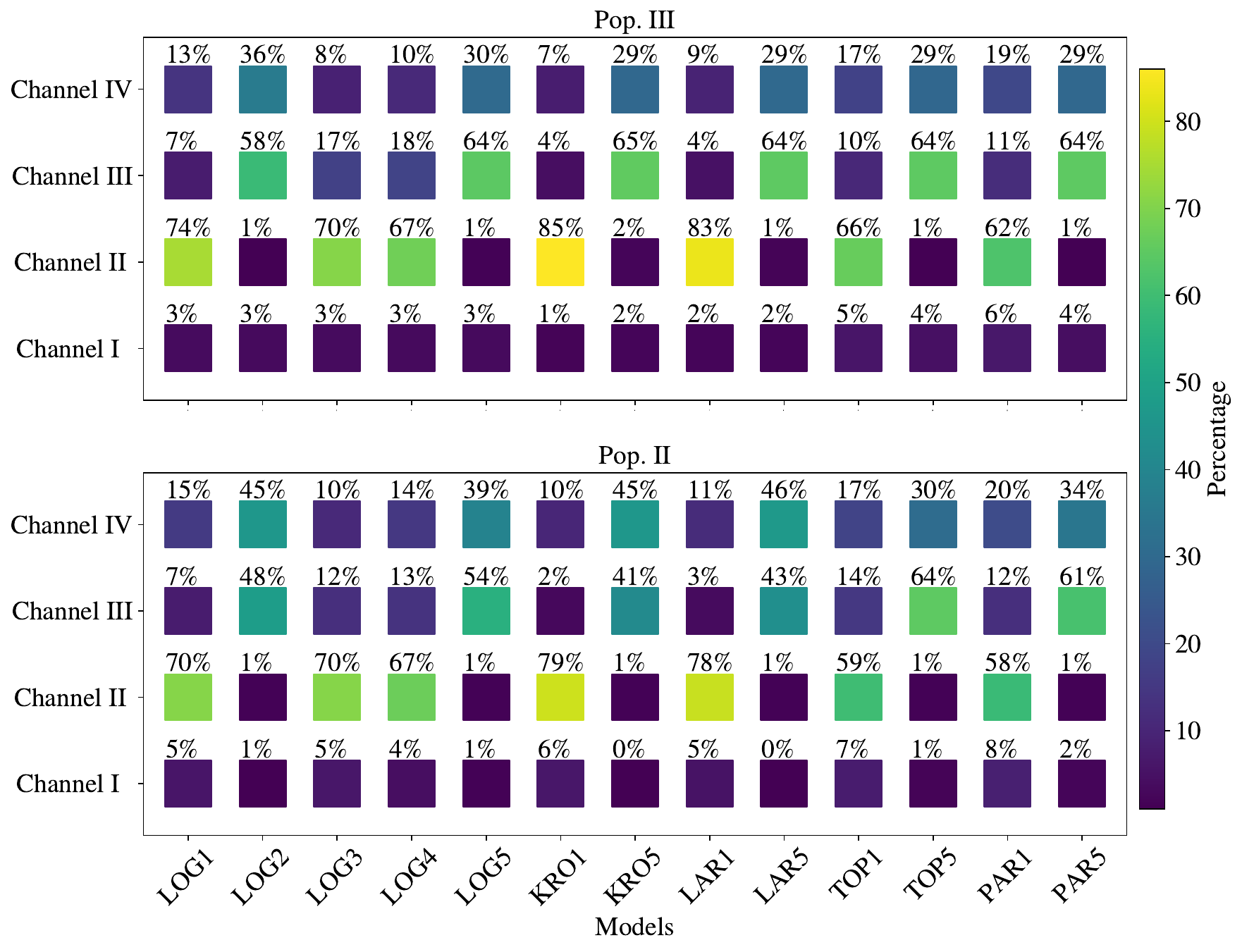} %
\caption{
    Percentage of BBH mergers that evolve via each of the four channels considered in this work. The x-axis refers to the simulation set, while the y-axis specifies the channel. The upper (lower) panel refers to BBHs that form from Pop.~III (II) stars. 
    }
    \label{fig:channel_percent}
\end{figure*}

The features of BBH mergers we described in the previous section (Section~\ref{sec:results_BSE}) can be interpreted by looking at the formation channels of our BBHs. Figure~\ref{fig:channel_percent} summarizes the main formation channels of BBH mergers from Pop.~III and Pop.~II stars. Tables~\ref{tab:perc_popIII} and \ref{tab:perc_popII} report the percentages in detail. 

We distinguish five main channels, following the definition by \cite{Iorio2023}. BBH mergers that go through Channel~0 do not undergo any mass transfer episodes during the evolution of their progenitors. Systems belonging to this channel are always very rare ($\ll{1}\%$). 
Since this channel is so uncommon, we do not show it in Figure~\ref{fig:channel_percent} and in the following Figures.

Channel~I is often referred to as the "traditional" formation channel of BBH mergers: the two progenitor stars undergo stable mass transfer before the formation of the first BH. Then, after the formation of the first BH, the system evolves through at least one common envelope.

In channel~II, the system evolves only via stable mass transfer episodes. Finally, in both channel~III and IV, the system undergoes at least one common envelope before the formation of the first BH. The only difference between channel~III and IV is that in the former the companion star preserves a residual of the original H-rich envelope at the time of the formation of the first BH, while in the latter the companion has  already been stripped of its envelope when the first BH forms.

\begin{table} 
    \caption{Percentage of BBH mergers from Pop.~III stars.    } 
    \begin{center}
        \begin{tabular}{rcccccc}
        \hline
Model & BBHm & Ch. 0 & Ch. I & Ch. II & Ch. III & Ch. IV \\
&  [\%] & [\%] & [\%] & [\%] & [\%] & [\%]  \\ 
\hline
LOG1 & 11.25 & 0.1 & 3.65 & 74.81 & 7.62 & 13.71 \\ 
LOG2 & 0.75 & 0.22 & 3.3 & 1.27 & 58.77 & 36.11 \\ 
LOG3 & 9.33 & 0.09 & 3.35 & 70.41 & 17.31 & 8.76 \\ 
LOG4 & 11.57 & 0.12 & 3.26 & 67.57 & 18.13 & 10.69 \\ 
LOG5 & 0.68 & 0.23 & 3.3 & 1.52 & 64.6 & 30.07 \\ 
KRO1 & 14.66 & 0.19 & 1.93 & 85.82 & 4.08 & 7.7 \\ 
KRO5 & 0.85 & 0.34 & 2.04 & 2.18 & 65.11 & 29.35 \\ 
LAR1 & 14.34 & 0.16 & 2.13 & 83.57 & 4.89 & 9.03 \\ 
LAR5 & 0.91 & 0.38 & 2.26 & 1.98 & 64.97 & 29.74 \\ 
TOP1 & 6.47 & 0.06 & 5.65 & 66.17 & 10.39 & 17.67 \\ 
TOP5 & 0.36 & 0.19 & 4.47 & 1.26 & 64.81 & 29.14 \\ 
PAR1 & 12.05 & 0.05 & 6.45 & 62.22 & 11.65 & 19.59 \\
PAR5 & 1.11 & 0.13 & 4.28 & 1.07 & 64.93 & 29.53 \\
        \hline
         \end{tabular}
   	\end{center}
\footnotesize{Column 1: Model; column 2: percentage of BBH mergers with respect to all simulated BBHs; columns 3, 4, 5, 6, and 7: BBH mergers formed via channel 0, I, II, III, and IV, respectively.}
 \label{tab:perc_popIII} 
 \end{table}
\begin{table} 
    \caption{Percentage of BBH mergers from Pop.~II stars.    } 
    \begin{center}
        \begin{tabular}{rcccccc}
        \hline
Model & BBHm & Ch. 0 & Ch. I & Ch. II & Ch. III & Ch. IV \\
&  [\%] & [\%] & [\%] & [\%] & [\%] & [\%]  \\ 
\hline
LOG1 & 13.53 & 0.02 & 5.94 & 70.48 & 7.59 & 15.44 \\ 
LOG2 & 0.97 & 0.12 & 1.3 & 1.44 & 48.11 & 45.88 \\ 
LOG3 & 10.88 & 0.03 & 5.98 & 70.55 & 12.34 & 10.71 \\ 
LOG4 & 14.46 & 0.02 & 4.03 & 67.06 & 13.37 & 14.79 \\ 
LOG5 & 0.86 & 0.11 & 1.48 & 1.53 & 54.96 & 39.05 \\ 
KRO1 & 16.23 & 0.03 & 6.16 & 79.64 & 2.85 & 10.01 \\ 
KRO5 & 1.15 & 0.23 & 0.79 & 1.66 & 41.21 & 45.93 \\ 
LAR1 & 16.15 & 0.04 & 5.28 & 78.93 & 3.37 & 11.35 \\ 
LAR5 & 1.16 & 0.22 & 0.91 & 1.6 & 43.11 & 46.57 \\ 
TOP1 & 8.52 & 0.02 & 7.42 & 59.61 & 14.73 & 17.99 \\ 
TOP5 & 0.48 & 0.1 & 1.9 & 1.44 & 64.8 & 30.76 \\ 
PAR1 & 15.13 & 0.01 & 8.53 & 58.46 & 12.52 & 20.32 \\
PAR5 & 1.36 & 0.06 & 2.01 & 1.48 & 61.36 & 34.33 \\
        \hline
         \end{tabular}
   	\end{center}
\footnotesize{Column 1: Model; column2: percentage of BBH mergers with respect to all simulated BBHs; columns 3, 4, 5, 6, and 7: BBH mergers formed via channel 0, I, II, III, and IV, respectively.}
 \label{tab:perc_popII} 
 \end{table}

\begin{figure}
\includegraphics[width=\columnwidth]{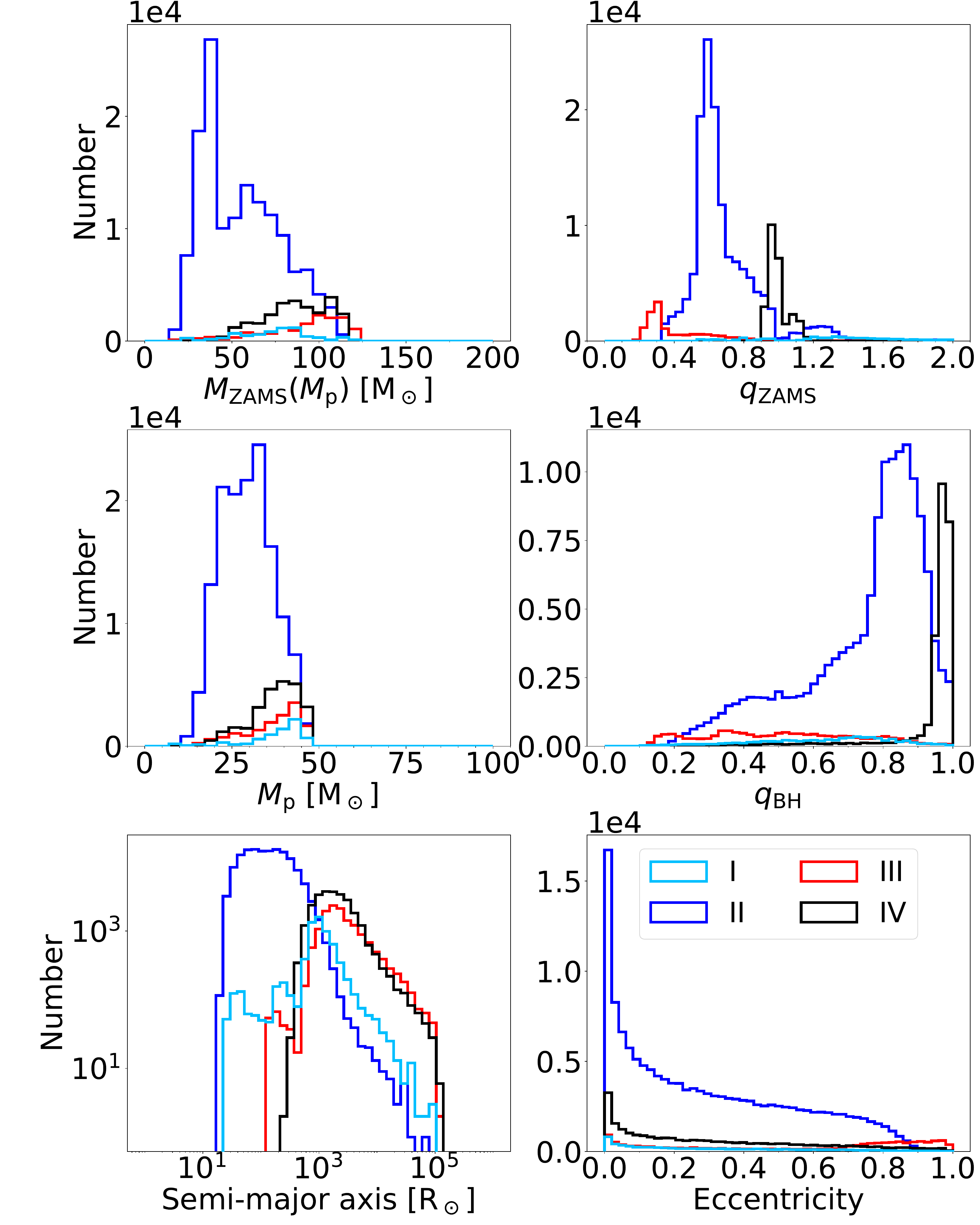}
    \caption{
    Main properties of BBH mergers and their progenitors in model LOG1  for Pop.~III stars. From top to bottom and from left to right: ZAMS mass of the progenitor of the primary BH $M_{\rm ZAMS}(M_{\rm p})$, mass ratio of the progenitors $q_{\rm ZAMS}=M_{\rm ZAMS}(M_{\rm s})/M_{\rm ZAMS}(M_{\rm p})$ (i.e. the ratio between the ZAMS mass of the progenitor of the secondary BH  and the ZAMS mass of the progenitor of the primary BH), mass of the primary BH ($M_{\rm p}$), mass ratio of the two BHs ($q_{\rm BH}=M_{\rm s}/M_{\rm p}$), initial semi-major axis ($a$), initial eccentricity ($e$). Light-blue line: channel I, blue line: channel II, red line: channel III, black line: channel IV.}
    \label{fig:LOG1_channels}
\end{figure}

\begin{figure}
\includegraphics[width=\columnwidth]{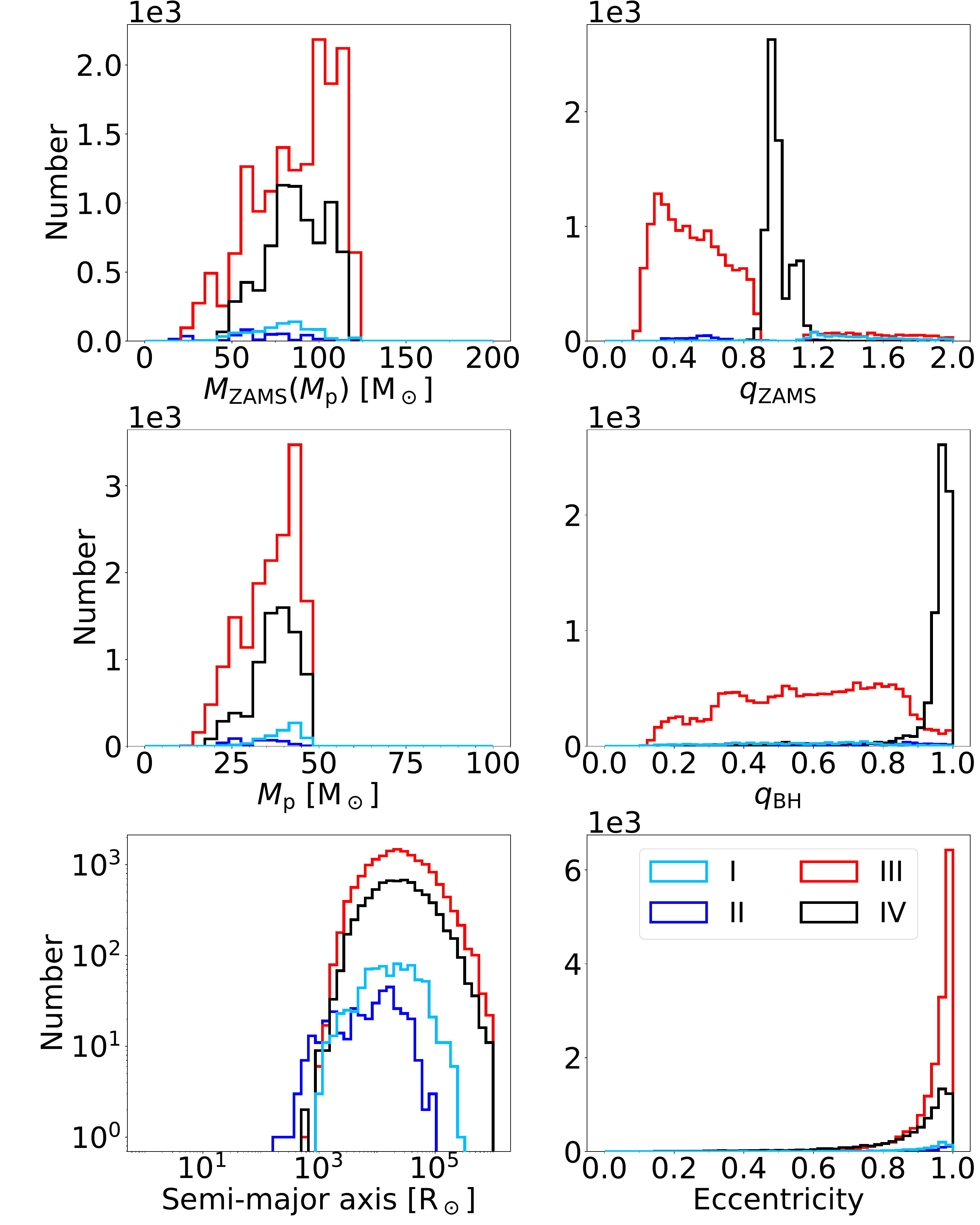} 
    \caption{
    Same as Figure~\ref{fig:LOG1_channels} but for model LOG5  for Pop.~III stars. }
    \label{fig:LOG5_channels}
\end{figure}

\begin{figure}
\includegraphics[width=\columnwidth]{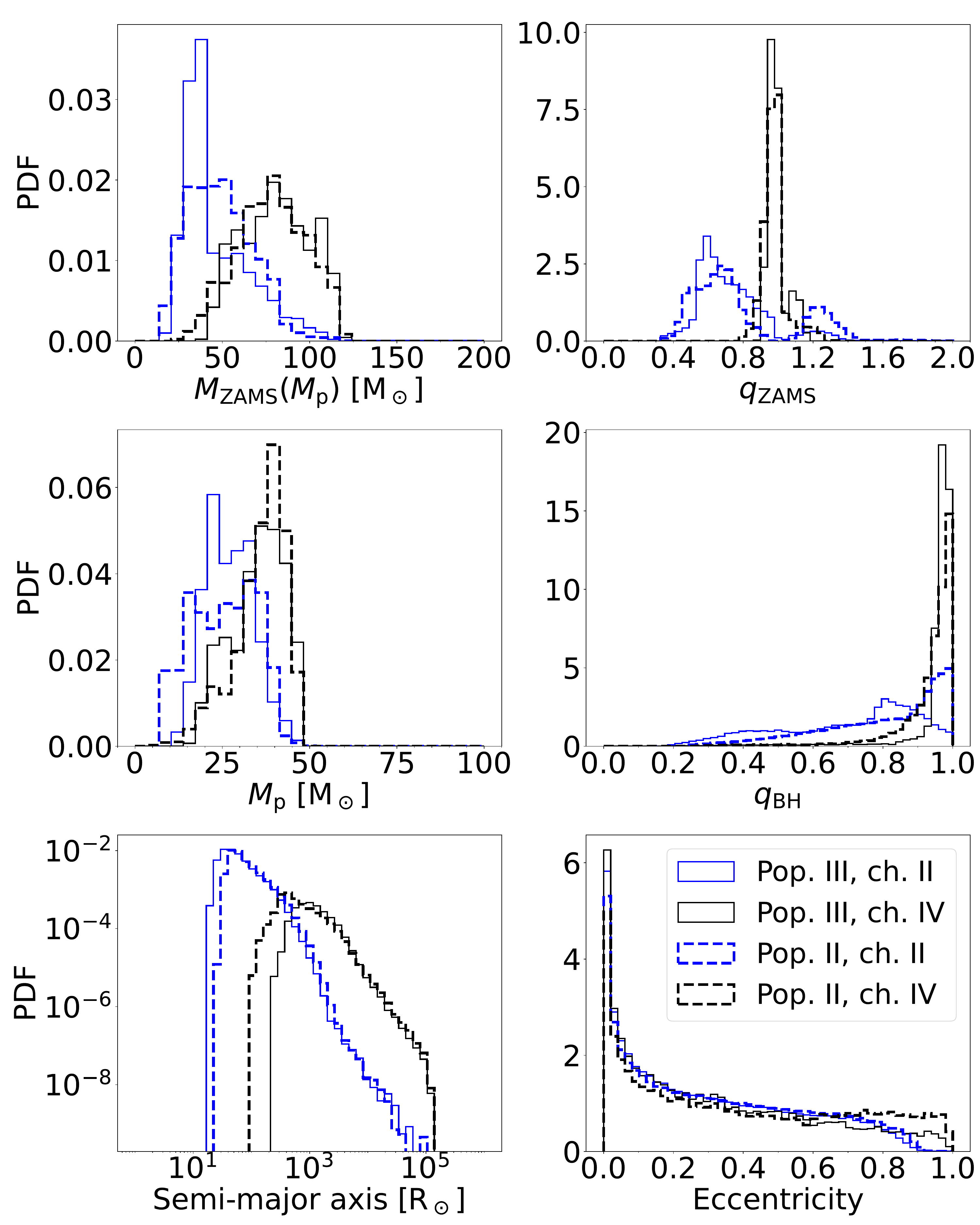}
    \caption{
    Same as Figure~\ref{fig:LOG1_channels} but here we show the KRO1 model and  compare BBHs from Pop.~III (solid  lines) and Pop.~II stars (dashed  lines). We show only channel~II and IV because they are the two most important channels (especially channel II, Fig.~\ref{fig:channel_percent}) and to make the plot more readable. }
    \label{fig:KRO1_channels}
\end{figure}

\begin{figure}
\includegraphics[width=\columnwidth]{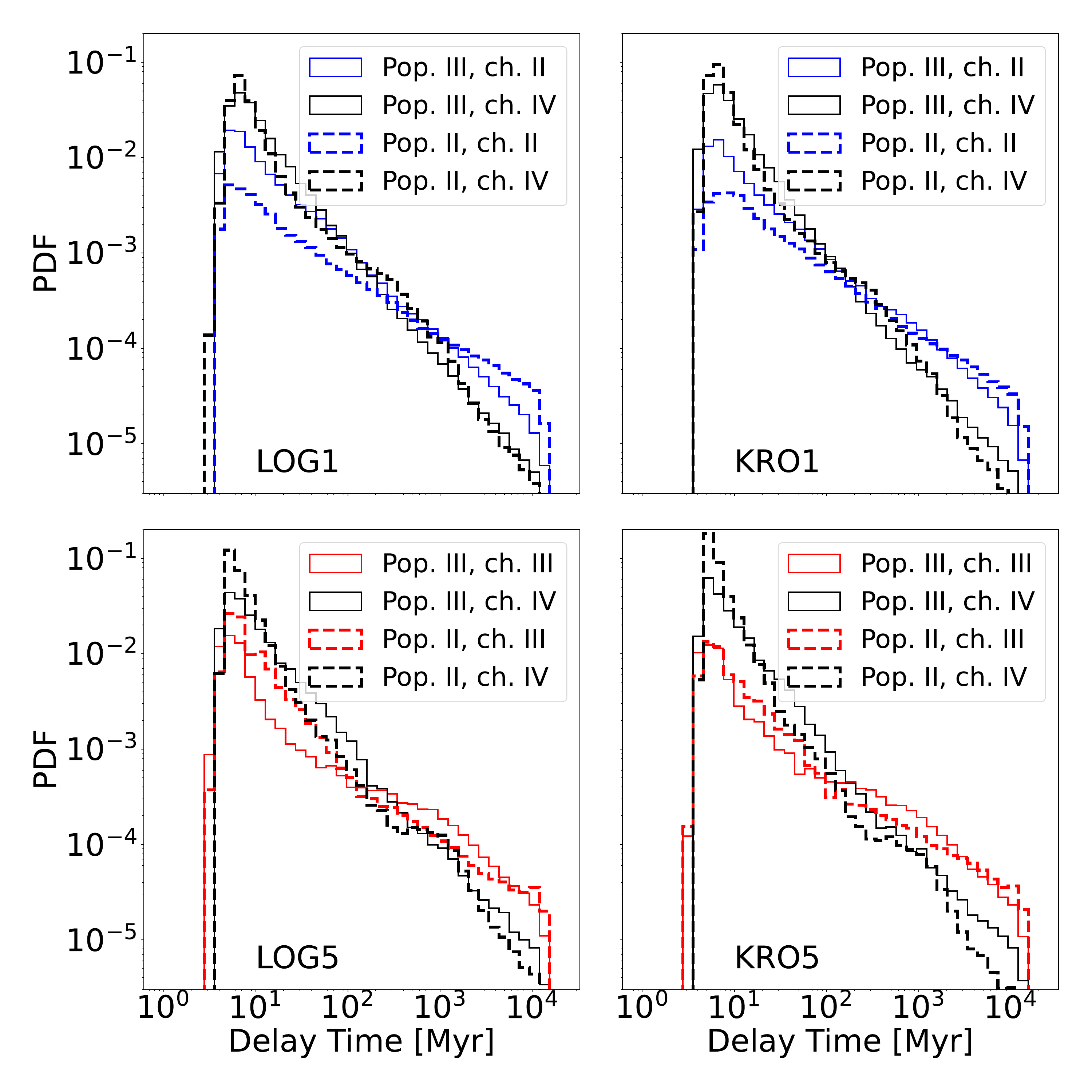} 
    \caption{
   Delay time distribution for Pop.~II and Pop.~III BBHs (dashed thick and solid thin lines, respectively). From left to right and from top to bottom: models LOG1, KRO1, LOG5 and KRO5. In each panel, we show only the two most important channels for each considered model:  channels II and IV in the upper panels, and channels III and IV in the lower panels. }
    \label{fig:tdelay_channels}
\end{figure}

The initial orbital period is the main driver of the relative differences among formation channels (Fig.~\ref{fig:channel_percent}). As we detail below, relatively short initial orbital periods (as in \citetalias{sana2012}) favour channel~II (i.e. stable mass transfer), while relatively long initial orbital periods (\citetalias{stacy2013}) favour channels~III and IV (i.e., formation channels with a common-envelope episode before the formation of the first BH). The main reason is that for short initial orbital periods the two progenitor stars undergo the first stable mass transfer episode early in their life (during the \ac{MS} or Hertzsprung-gap phase), while for large initial orbital periods the first interaction happens in a late evolutionary phase, when the primary star has developed a large radius and a well-defined core structure. This result holds for both Pop.~III and Pop.~II stars (Fig.~\ref{fig:channel_percent}).

Figures~\ref{fig:LOG1_channels} and \ref{fig:LOG5_channels}
show the behaviour of Pop.~III BBHs and their progenitor stars in models LOG1 and LOG5, respectively. 
We  show only models LOG1 and LOG5 for the sake of brevity: Models LOG3, LOG4, KRO1, LAR1,  TOP1 and PAR1   behave in a similar way to LOG1 with respect to the formation channels (these are the models that adopt the initial \citetalias{sana2012} orbital period distribution), while models  LOG2, KRO5, LAR5, TOP5, and PAR5 behave in a similar way to  LOG5, which adopts the initial \citetalias{stacy2013} orbital period distribution.

Channel~II (i.e. stable mass transfer) is the dominant channel for model LOG1 and for all the other models that adopt the initial \citetalias{sana2012} distribution of the orbital periods. Figure~\ref{fig:LOG1_channels} shows that most channel~II systems are associated with short initial semi-major axis $a=10-10^3$ R$_\odot$. These short initial semi-major axes are heavily suppressed with the orbital period distribution by \citetalias{stacy2013} (Fig.~\ref{fig:LOG5_channels}).

Figure~\ref{fig:LOG1_channels} also shows that channel~II is associated with relatively low mass ratios in the ZAMS ($q_{\rm ZAMS}\sim{0.5-0.9}$) and relatively low mass ratios between the two final BHs ($q_{\rm BH}=0.75-0.9$) for Pop.~III stars.
Hence, the predominance of channel~II in the models adopting the \citetalias{sana2012} orbital period distribution explains why these models have a preference for BBH mass ratios $q_{\rm BH}=0.75-0.9$ in the case of Pop.~III stars, as discussed in the previous Section (Fig.~\ref{fig:q}). In these systems, the mass difference between primary and secondary stars is sufficiently large that the system undergoes the first mass transfer while the secondary is still on the \ac{MS}.

In contrast, channels~III and IV are the dominant channels for all the models that adopt the \citetalias{stacy2013} orbital period distribution. As shown by, e.g., Fig.~\ref{fig:LOG5_channels}, the large initial semi-major axes of distribution \citetalias{stacy2013} suppress systems with initial orbital separation $a<10^3$ R$_\odot$, hence suppressing channel~II. Channel~III and IV preferentially arise when $a\sim{10^3-10^5}$ R$_\odot$. In this case, mass transfer takes place only when the radii of the two stars become very large, i.e. in the late evolutionary stages. Channel~IV is the preferred channel of equal-mass stars, that evolve nearly at the same time and strip off each other's envelopes. It mainly leads to the formation of equal-mass BBHs, explaining the preference of these models for equal-mass mergers (Fig.~\ref{fig:q}). In contrast, channel~III has a preference for markedly unequal-mass systems, explaining the population of BBHs with $q\leq{}0.6$ in models LOG5, KRO5, LAR5, TOP5, and PAR5 (especially for Pop.~II stars, Fig.~\ref{fig:q}).

Figure~\ref{fig:KRO1_channels} compares the properties of BBH mergers from Pop.~III and Pop.~II stars in the case of model KRO1 (the fiducial model for Pop.~II stars). Since channels I and III are less important than channels II and IV in this model, we show only the latter channels for simplicity. This Figure shows that Pop.~III and Pop.~II stars have a  very similar behaviour in the case of channel IV. As for channel~II, we see three main differences: Pop.~II stars have a preference for i)  higher ZAMS mass $M_{\rm ZAMS}(M_{\rm p})$, ii) larger mass ratios in the ZAMS $q_{\rm ZAMS}$, and iii) and larger BBH mass ratios  $q_{\rm BH}$ with respect to Pop.~III stars. The correlation between these three properties explains why Pop.~II stars tend to produce equal-mass BBHs,  while Pop.~III stars produce BBHs with a mass ratio peaking at $q_{\rm BH}\approx{0.8}$  (Fig.~\ref{fig:q}). 
 
Figure~\ref{fig:tdelay_channels} compares the delay time distribution of Pop.~III and Pop.~II binary systems if we consider models KRO1 (fiducial model for Pop.~II), LOG1 (fiducial model for Pop.~III), LOG5 and KRO5. Channel~IV is skewed toward the shortest delay times, for both Pop.~III and II binary stars, because it is associated with the most efficient orbital shrinking during common envelope. The shrinking is more efficient for Pop.~II stars (especially in models LOG5 and KRO5), because they reach even larger radii in their late evolutionary stages (see Figures~\ref{fig:HR} and \ref{fig:RvsAge}). This explains why the overall delay time distribution of Pop.~II systems (Figure~\ref{fig:delay_time}) has an excess at very low values ($t_{\rm del}\leq{10}$ Myr) in models LOG2, LOG5, KRO5, LAR5, TOP5, and PAR5. In contrast, channel~II is associated with long delay times, because stable mass transfer is not as efficient as common envelope in reducing the orbital separation. 

Tables~\ref{tab:perc_popIII} and \ref{tab:perc_popII}, and Figure~\ref{fig:channel_percent} show that both Pop.~III and Pop.~II stars behave in a very different way from more metal rich binary systems. In fact, only $\lesssim{5.7}\%$ ($\lesssim{7.5}\%$) of all BBH mergers evolve via channel~I  in the case of Pop.~III (II) binary stars. For comparison, \cite{Iorio2023} show that between 50 and 80\% of all BBH mergers evolve via channel~I at metallicity between $Z=2\times{}10^{-3}$ and $Z=10^{-2}$ (see Fig.~14 of \citealt{Iorio2023} for $\alpha=1$). 

The abundance of channel~II systems and the dearth of channel~I systems for Pop.~III and II binary stars  with respect to metal-rich  binary systems ($Z\in{}[2\times{}10^{-3}-10^{-2}]$) are a consequence of the large BH masses at such low metallicity. In fact, both channel~I and II systems go through a stable mass transfer before the formation of the first compact object, and then undergo a second mass transfer after the formation of the first compact object. The only difference between the two channels is that the mass transfer episode between the companion star and the first-born BH becomes unstable in channel I and remains stable in channel II. In \sevn{} (as in most binary population-synthesis codes) the stability of mass transfer is evaluated through a critical mass ratio $q_{\rm crit}$ between the donor and the accretor: the systems with mass ratio $q\geq{}q_{\rm crit}$ undergo a dynamically unstable mass transfer (i.e., a common-envelope episode), while mass transfer remains stable in the other cases \citep{hurley2002}. Since most our Pop.~II and III BHs in tight binary systems are relatively massive ($\gtrsim{}20$ M$_\odot$, Fig.~\ref{fig:MassSpec}), we have that $q<q_{\rm crit}$ in most binary systems (where $q$ is the mass ratio between the donor star and the first born BH), ensuring the stability of most mass-transfer episodes. 

Furthermore, the abundance of channel~II versus channel~I systems depends on the assumed value of common-envelope efficiency $\alpha_{\rm CE}$. We have re-run the fiducial case LOG1 with $\alpha_{\rm CE}=3$. We find that for Pop.~III stars, the percentage of channel~II systems decreases from 75\% for $\alpha_{\rm CE}=1$ down to 51\% for $\alpha_{\rm CE}=3$. This happens because, when $\alpha_{\rm CE}$ is large, the common envelope process is less efficient in shrinking the orbital separation. In contrast, the relative abundance between channel~I and II systems is not significantly affected by our assumption that mass transfer is always stable for \ac{MS} and Hertzsprung-gap donors. In fact, both channels~I and II undergo a stable mass transfer when the primary star is still a \ac{MS} or an Hertzsprung-gap star. Relaxing the aforementioned assumption has a more sizable impact on the evolution of channels~III and IV systems, which evolve via common envelope before the formation of the first-born BH.

\subsection{BBH mergers with primary above the mass gap}

\begin{figure}
\includegraphics[width=\columnwidth]{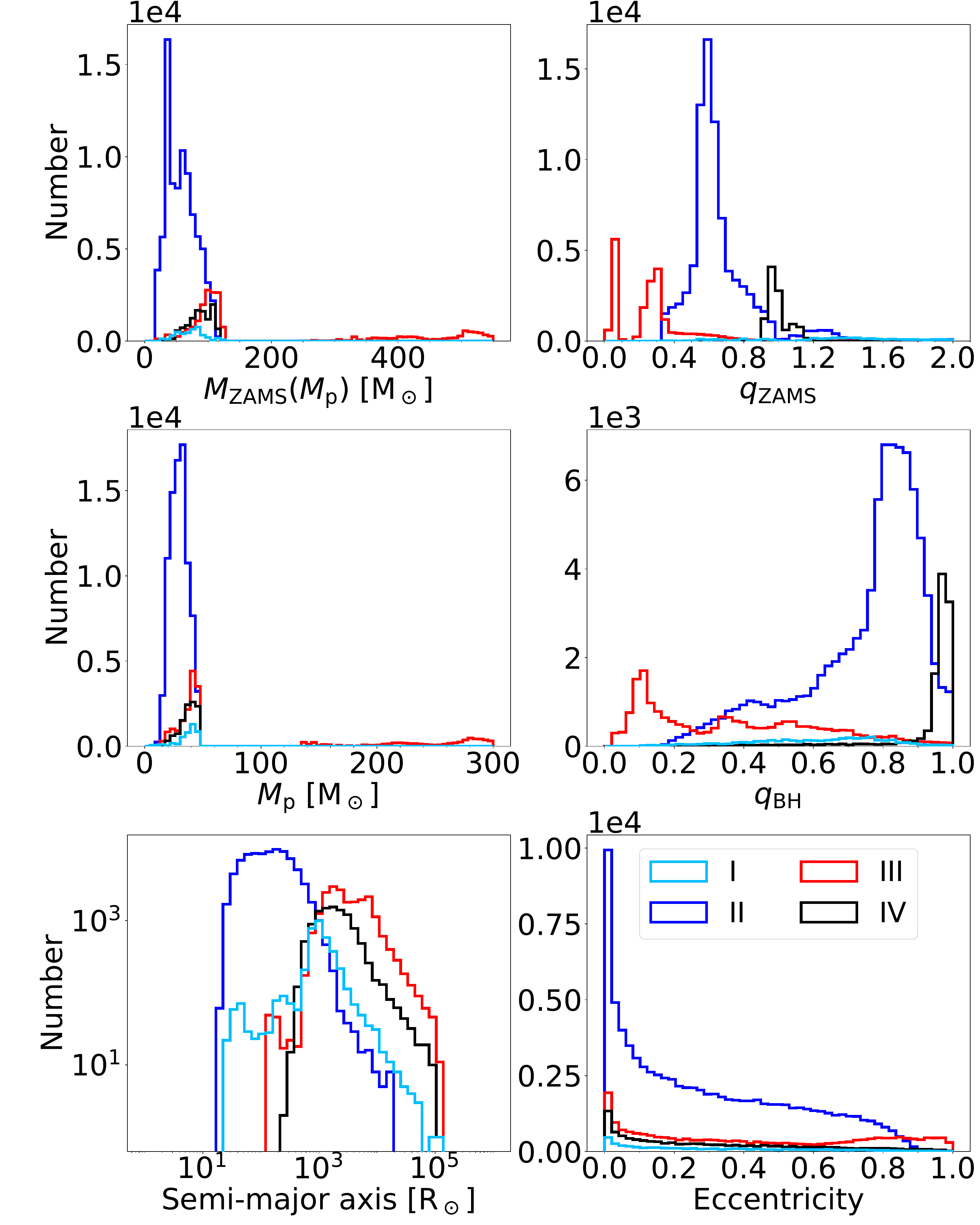}
    \caption{
    Same as Fig.~\ref{fig:LOG1_channels} but for model LOG3 for Pop.~III stars. }
    \label{fig:LOG3_channels}
\end{figure}

Channel~III is the key to interpret the abundance of BBHs with primary mass above the gap in model~LOG3. These systems are associated with a population of binary stars with very low $q_{\rm ZAMS}$ (Figure~\ref{fig:LOG3_channels}), mostly following the relation $q\approx23\,{}M_\odot/M_\mathrm{ZAMS,1}$ with $M_\mathrm{ZAMS,1}\gtrsim250 \ \Msun$  and large initial separations $a\gtrsim10^3 \ \Msun$. Such systems evolve through channel III triggering a Roche-lobe overflow episode that becomes unstable when the primary star enters the red super-giant phase. Systems with smaller initial separation merge due to a double Roche-lobe overflow. 
Only systems with initial $q_{\rm ZAMS}<0.15$ can evolve through this channel. This is the reason why most of the high-mass primaries come from the model LOG3, followed by LOG4 and LOG1, LOG2 (see Figs.~\ref{fig:IC_IMF}, \ref{fig:IC_q} and \ref{fig:IC_P_e}). 
The fact we do not see this feature in the model LOG3 of Pop.~II stars is related to the evolution of radius. High-mass Pop II stars reach the radius to start the interaction earlier than Pop. III stars (see Fig.~\ref{fig:RvsAge}).

\subsection{Comparison with previous work}

Several authors have explored the formation of BBHs from Pop.~III stars \citep[e.g.,][]{Kinugawa2014,Kinugawa2016,Kinugawa2020,Hartwig2016,belczynski2017,liubromm2020GW190521,tanikawa2021,tanikawa2022,tanikawa2022b,wang2022}. The ZAMS mass range we consider in this work is comparable to the one explored by  \cite{tanikawa2021}. In our models, we do not have any mergers with both BHs above the mass gap, while these are very common in their fiducial model. This discrepancy stems from the intrinsic differences between our single star evolution models. In fact, the very massive stars ($M_{\rm ZAMS}>200$ M$_\odot$) considered by \cite{tanikawa2021} end the \ac{MS} as compact blue super-giant stars, while our very massive stars  expand  during the \ac{MS} and become red super-giant stars already at the end of the \ac{MS}. As a consequence, the very massive binary systems by \cite{tanikawa2021} undergo stable mass transfer and leave BHs above the mass gap, while our  very massive binary systems start an unstable  common envelope phase as soon as they leave the \ac{MS} and merge prematurely, before becoming BHs.
 
The same line of reasoning explains why our delay times (Figs.~\ref{fig:delay_time} and Figs.~\ref{fig:tdelay_channels}) are generally much shorter than the one presented, e.g., in Figure~3 from  \cite{tanikawaGW190521}. Almost all BBH mergers from Pop.~III stars evolve via stable mass transfer in the models by \cite{tanikawaGW190521} and thus have long delay times, while our channel~IV mergers (which do not form in \citealt{tanikawaGW190521}) have very short delay times. These results confirm the key role of single star evolution (including uncertainties about core overshooting, convection and rotation) for the formation of merging BBHs.

\section{Summary and conclusions}
\label{sec:summary}

We have presented a new set of Pop.~III stars ($Z=10^{-11}$) obtained with the  stellar evolution code {\sc parsec} \citep{Bressan2012,Costa2019,costa2021}. Our Pop.~III stars range from 2 to 600 M$_\odot$. With respect to Pop.~II stars ($Z=10^{-4}$), Pop.~III stars with initial mass $M_{\rm ZAMS}\in{[14,40]}$ M$_\odot$ evolve with much more compact radii ($R\ll{}10^2$~R$_\odot$).
Furthermore, the most massive Pop.~III stars ($\Mz{}{} > {100}$ M$_\odot$) end their lives as blue super-giant stars, whereas Pop.~II stars in the same mass range die as yellow and red super-giant stars.

We use these tracks as input tables for our fast binary population synthesis code \sevn{} \citep{spera2019, mapelli2020, Iorio2023}, in order to study the population of BHs and BBHs born from Pop.~II and Pop.~III stars. We explore a variety of initial conditions for our binary stars, including a flat-in-log, a Kroupa (\citetalias{kroupa2001}), a Larson (\citetalias{larson1998}), a top-heavy, and a Park (\citetalias{Park2023}) IMF.

We estimate similar BH masses from the evolution of single massive Pop.~II and Pop.~III stars. In our fiducial model, the maximum BH mass below the pair-instability gap is 91 and 86 M$_\odot$ for Pop.~II and III stars, respectively. Above the gap, both Pop.~II and Pop.~III stars produce BHs more massive than $\approx{230}$ M$_\odot$ if they can achieve a ZAMS mass of $\gtrsim{240}$ M$_\odot$ (Figure~\ref{fig:MassSpec}). Assumptions on core overshooting, envelope undershooting, and stellar rotation can significantly affect this result, because they influence the mass of the He and CO cores, hence the central temperature and density. Furthermore, these results stem from the assumption that the residual H-rich envelope is not ejected during the failed supernova \citep[e.g.,][]{Costa2022}.

Most BBH mergers from both Pop.~II and  Pop.~III stars have  primary BH mass below the mass gap. In order to populate the region above the gap, we need very compact stellar radii, that can be achieved either with fast rotation (chemically homogeneous evolution, \citealt{deMink2015}) or by suppressing core overshooting \citep{tanikawa2022b}. With our evolutionary models, we find no mergers with secondary BH mass above the gap. We expect that only dynamics of dense stellar systems can pair up BHs with both primary and secondary mass above the gap, and populate the gap as well \citep[e.g.,][]{Mapelli2022,wang2022}.

The mass ratios are one of the main signatures of Pop.~III versus Pop.~II BBHs. In most of our models, Pop.~II BBHs are predominantly equal-mass systems, whereas Pop.~III BBHs have a peak at mass ratio $q_{\rm BH}=0.8-0.9$. This difference is too subtle for current detectors, even at a population level, but it can be investigated with next-generation ground-based detectors.

A distinctive signature of Pop.~III and II BBHs with respect to BBHs born from metal-rich stars are the evolutionary channels. Assuming the orbital period distribution from \cite{sana2012}, the vast majority ($60-80$\%) of Pop.~III and II progenitor stars of BBH mergers evolve via channel~II, i.e. just stable mass transfer, with no common envelope. 
In contrast, at higher metallicity ($Z\sim{2\times{}10^{-3}-10^{-2}}$) and with the same set-up for binary evolution, the dominant evolutionary pathway ($50-80$\% BBH mergers) becomes channel I, characterized by a common envelope between the first-born BH and its companion star \citep{Iorio2023}. 

If we instead assume that Pop.~III binary systems have longer orbital periods (e.g., \citealt{stacy2013}), both channels~I and II are suppressed: most Pop.~III and Pop.~II BBHs form from an early common-envelope episode that involves the two progenitor stars, before the formation of the first-born BH (channels III and IV).

Overall, our models show that Pop.~III and II stars produce a similar BBH population, especially if we adopt the same IMF and initial orbital properties. 
The actual IMF and maximum mass of metal-poor and metal-free stars are two of the main sources of uncertainty.


\section*{Acknowledgements}

We are grateful to  Simon Glover, Tilman Hartwig, Tomoya Kinugawa, Mario Spera, and Ataru Tanikawa 
for their enlightening comments.  GC, GI, MM, and FS acknowledge financial support from the European Research 
Council for the ERC Consolidator grant DEMOBLACK, under contract no. 770017.
MM and RSK acknowledge financial support from the German Excellence Strategy via the Heidelberg Cluster of Excellence (EXC 2181 - 390900948) ``STRUCTURES''.
This research made use of \textsc{NumPy} \citep{Harris20}, \textsc{SciPy} 
\citep{SciPy2020}, \textsc{IPython} \citep{Ipython}. For the plots we used \textsc{Matplotlib} \citep{Hunter2007}.

\section*{Data Availability}

The data underlying this article and the \sevn{} configurations files are available on Zenodo at \url{https://doi.org/10.5281/zenodo.7736309} \citep{Costa2023_zenodo}.
\sevn{} is publicly available at \url{https://gitlab.com/sevncodes/sevn.git}: the version used in this work is the commit \texttt{0f9ae3bf}  in the branch \texttt{Costa23popIII}   (\url{https://gitlab.com/sevncodes/sevn/-/tree/Costa23popIII}).
Further data will be shared on reasonable request to the corresponding authors. 
  


\bibliographystyle{mnras}
\bibliography{biblio} 

\end{document}